\documentclass[aps,twocolumn,letterpaper,longbibliography]{revtex4-1}

\usepackage{amssymb}
\usepackage{amsmath}
\usepackage{graphicx}
\usepackage{extarrows}
\usepackage{url}
\usepackage{varioref}
\usepackage{hyperref}
\usepackage{cleveref}
\usepackage{color}
\usepackage{changes}
\usepackage{float}

\newcommand\ket[1]{\ensuremath{|#1\rangle}}
\newcommand\bra[1]{\ensuremath{\langle#1|}}
\newcommand\mean[1]{\ensuremath{\left<#1\right>}}
\newcommand{\vk}{\mathbf{k}}
\newcommand{\vvr}{\mathbf{r}}

\begin{document}
\title{Quantum Simulation of Coherent Hawking-Unruh Radiation}
\author{Jiazhong Hu}
\author{Lei Feng}
\author{Zhendong Zhang}
\author{Cheng Chin}
\affiliation{James Franck Institute, Enrico Fermi Institute and Department of Physics, University of Chicago, Chicago, Illinois 60637, USA}

\begin{abstract}
Exploring quantum phenomena in a curved spacetime is an emerging interdisciplinary area relating many fields in physics such as general relativity \cite{Hawking1974,Hawking1975,Unruh1976,Wald1994}, thermodynamics \cite{Bekenstein1973,Hawking1976,Wald1994}, and quantum information \cite{Hayden2007,Giddings2013}.
One famous prediction is the Hawking-Unruh thermal radiation \cite{Unruh1976}, the manifestation of Minkowski vacuum in an accelerating reference frame.
We simulate  the radiation 
by evolving a parametrically driven Bose-Einstein condensate of $\approx 10^5$ atoms \cite{Logan2017}, which radiates coherent pairs of atoms with opposite momenta.
We observe a matterwave field which follows a Boltzmann distribution for a local observer.
The extracted temperature and entropy from the atomic distribution are in agreement with Unruh's predictions \cite{Unruh1976}.
We further observe the long-distance phase coherence and temporal reversibility of emitted matter-waves, hallmarks that distinguish Unruh radiations from classical counterparts.
Our results may lead to further insights regarding the nature of the Hawking and Unruh effects and behaviors of quantum physics in a curved spacetime.
\end{abstract}

\maketitle

Applying quantum mechanics to gravitational systems is one of the hot areas to explore the not-yet-understood physics of quantum gravity. Ideas such as Hawking radiation \cite{Hawking1974,Hawking1975}, gauge-gravity duality \cite{Maldacena1998}, and the black hole information paradox \cite{Hawking1976v2,Susskind2006,Almheiri2013} inspire understanding of the role of quantum mechanics in gravitational fields, and are essential steps toward a new approach to the foundations of physics.

Among these pioneering approaches, Unruh radiation \cite{Unruh1976} is predicted to describe quantum fluctuations in a non-inertial frame.
A vacuum state of fields in the Minkowski space can appear as a thermal state to an accelerating observer. The thermal radiation is characterized by the Unruh temperature $T_\mathrm{U}$ \cite{Unruh1976}, and this temperature depends on the acceleration of the observer $A$ as
\begin{equation}
T_\mathrm{U}={\hbar A\over 2\pi c k_\mathrm{B}},\label{eq1}
\end{equation}
where $k_\mathrm{B}$ is the Boltzmann constant, $\hbar$ is the reduced Planck constant and $c$ is the speed of light. Because of the equivalence of inertial and gravitational acceleration, this surprising phenomenon shares the same root as the Hawking radiation \cite{Hawking1975} near the black hole horizon.
Thus the Unruh radiation is also known as Hawking-Unruh radiation. Experimentally, it is extremely challenging to observe Unruh effect; an enormous acceleration of $A=2.5\times 10^{14}$~m/s$^2$ is required to create an Unruh radiation of $T_\mathrm{U}=~1~\mu$K.

\begin{figure}[!htb]
\begin{center}
\includegraphics[width=87mm]{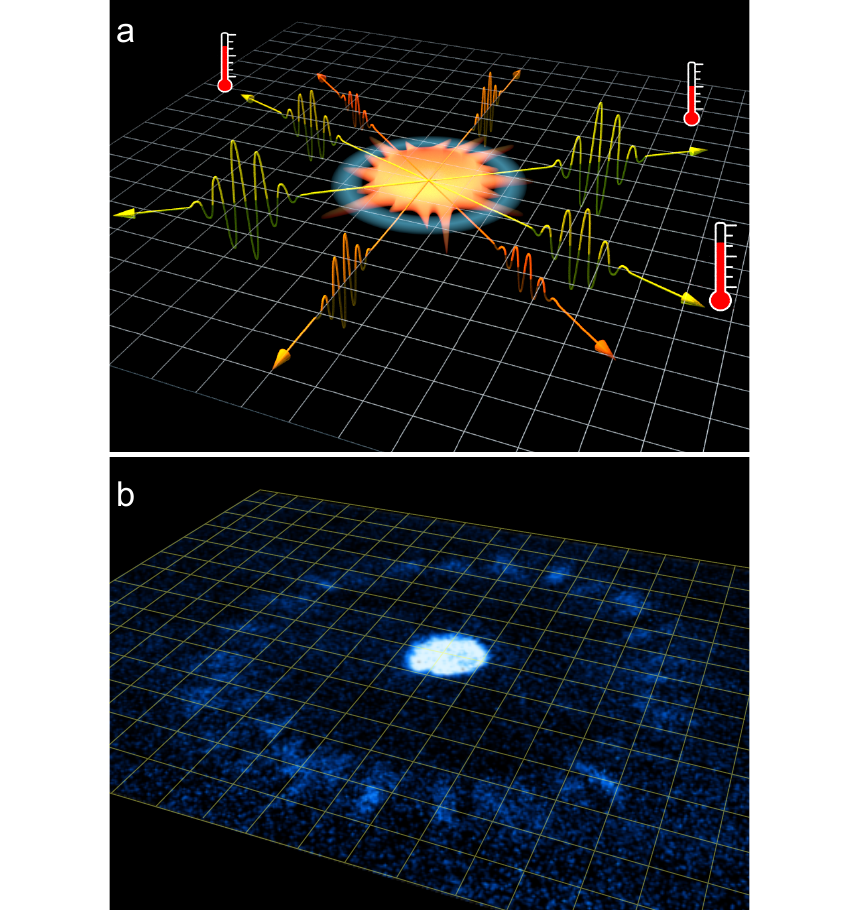}
\caption{\textbf{Quantum simulation of Hawking-Unruh radiation.} \textbf{a}, To an accelerating observer, a vacuum state in the inertial frame appears identical to a thermal state. \textbf{b}, We simulate the Hawking-Unruh effect by a pair-creation process in a driven condensate, whose evolution is equivalent to a coordinate transformation to an accelerating frame.
The matter-wave field shares the same characteristics as the Unruh radiation: it is locally indistinguishable from a Boltzmann distribution, but is long-range coherent and temporally reversible.} \label{fig1}
\end{center}
\end{figure}

Consider a quantum field $\Psi_0$ in an inertial frame which is transformed into a new field $\Psi_\mathrm{R}$ according to an accelerating observer. Such conversion is realized by the Rindler transformation $\Psi_\mathrm{R}=\hat R_A \Psi_0$ \cite{Wald1994}, where $\hat R_A$ is the operator which maps the quantum states to the accelerating basis. 

We propose that the frame transformation $\hat R_A$ can be simulated based on an evolution operator $\hat U(\tau)=e^{-i\mathcal H \tau/\hbar}$ such that (see Fig.~1)
\begin{equation}
\hat R_A\Psi_0=\hat U(\tau)\Psi_0,
\end{equation}
where the time $\tau$ in the lab frame acts as a parameter to control the acceleration $A$ in the simulated frame. The Hamiltonian given by $\mathcal H=i\hbar\sum_k g_k(a^\dagger_k a^\dagger_{-k}-a_k a_{-k})$ generates the frame boost, where $a_k$ ($a^\dagger_k$) is the annihilation (creation) operator of a particle with momentum $k$ and $g_k$ is the coupling constant. We can thus emulate the physics in the accelerating frame based on a bench-top experiment without the need to greatly accelerate the sample. (See Ref.~ \cite{Su2016} and Methods for details.)

In this paper we demonstrate quantum simulation of Unruh effect by only considering the momentum modes with the same amplitude $|\vec k|=$ constant (see Fig.~1). Modulating the interactions between condensed atoms, we engineer the Hamiltonian $H$, which  approximates (see Methods)
\begin{equation}
H=i\hbar g\sum_{|\vec k|=k_f}(a^\dagger_k a^\dagger_{-k}-a_k a_{-k}),\label{3}
\end{equation}
where $k_f$ is a constant. Given this Hamiltonian $H$, 
a condensate acts as a vacuum that radiates atoms to about $300$ momentum modes, sufficient to build statistics, test the Boltzmann distribution, and extract the effective temperature $T$ of the matter-waves fields. To verify 
that our system simulates the Unruh physics, we demonstrate the spatial coherence and reversibility of the matter-waves fields, which clearly distinguishes Unruh radiation from classical counterparts.

The connection between the dynamics of our system and the Rindler frame transformation with acceleration $A$ can be best understood from the evolution of the bosonic fields
\cite{Su2016},
\begin{gather}
 \begin{bmatrix} a_k(\tau) \\ a^\dagger_{-k}(\tau) \end{bmatrix}
 =\hat{\mathcal R}
   \begin{bmatrix} a_k(0) \\ a^\dagger_{-k}(0) \end{bmatrix},
\end{gather}
where $\hat{\mathcal R}= e^{g\tau\sigma_x}$ is the Rindler coordinate transformation, $\sigma_x$ is the $x-$component of the Pauli matrix. The acceleration $A$ in the simulated accelerating frame is given by (see Methods)
\begin{equation}
A={2\pi c E_{k_f} \over\hbar\ln(1+1/\bar n)}\xrightarrow[\bar n \gg1] {}   {2\pi c E_{k_f}   \over \hbar}\bar n, \label{eq4}
\end{equation}
where $E_{k_f}$ is the energy of the excitation with momentum $k=k_f$ and $\bar n(\tau)=\sinh^2(g\tau)$ is the mean population in one momentum mode (see Methods). In the large population limit $\bar{n}>>1$, the acceleration scales linearly with $\bar{n}$.

\begin{figure}
\begin{center}
\includegraphics[width=87mm]{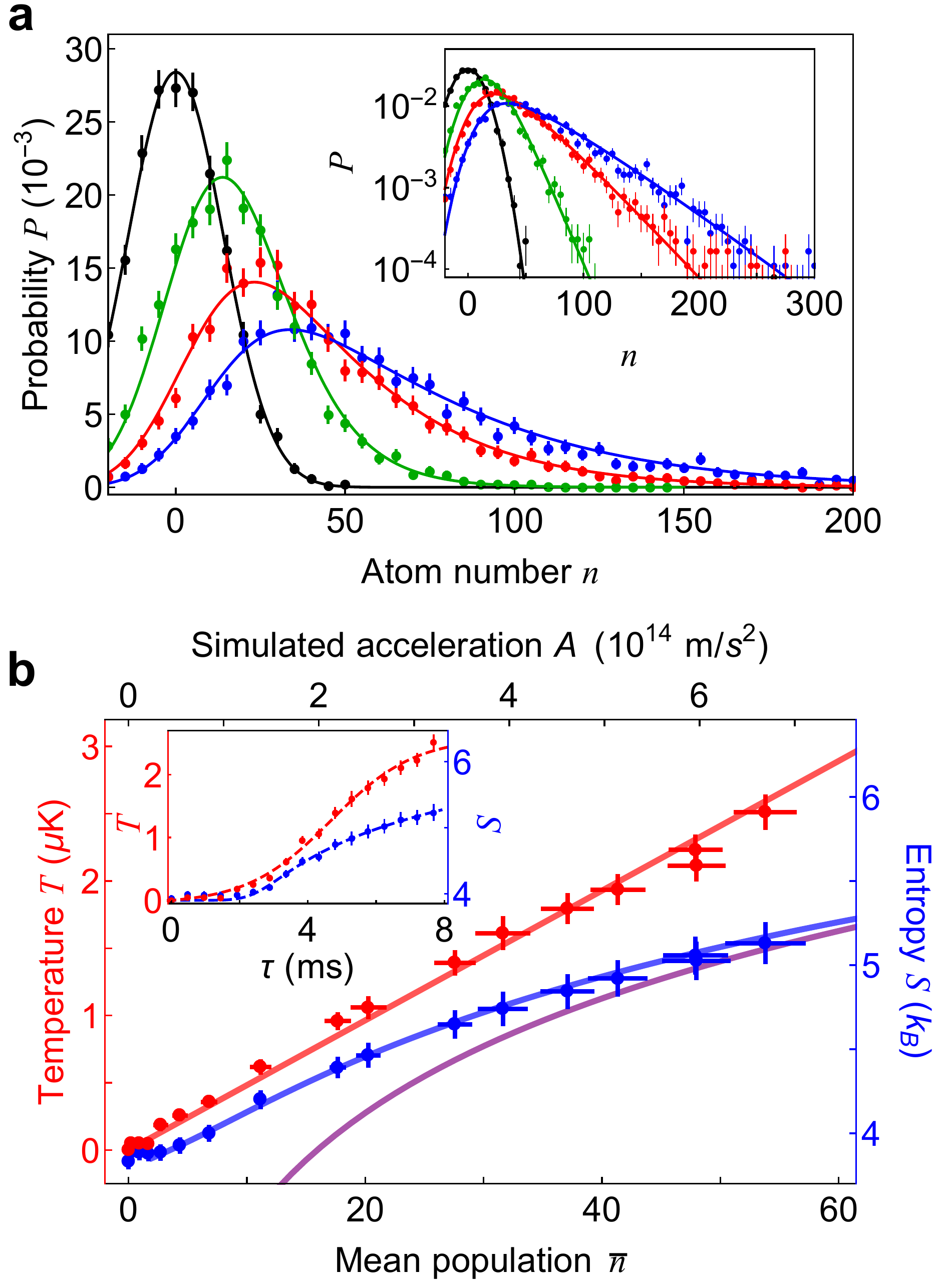}
\caption{\textbf{Thermal behavior of the matter-wave emission.}
\textbf{a} shows the measured probability distribution $P(n)$ within a 2$^\circ$ slice of the emission pattern after modulation time $\tau=~$0, 3.36, 4.8 and 6.24~ms (black, green, red and blue circles). The solid lines are fits based on a thermal model (see Methods). The inset shows the data in the log scale. \textbf{b} shows the effective temperature $T$ (red circles) and entropy per mode $S$ (blue circles) versus the mean population per mode. The derived acceleration $A$ is shown on the top. The red solid line is a fit of $T=\kappa A/c$. The blue solid line is the prediction that includes the detection noise while the purple line is the prediction excluding the noise. The inset shows the evolution of $T$ and $S$. The dashed lines are guides to the eye.
Here the condensate's radius is 13~$\mu$m. The scattering length is modulated at frequency $\omega/2\pi~=~2.1$~kHz with a small offset of $a_{dc}~=~3a_0$ and an amplitude of $a_{ac}~=~50a_0$, where $a_0$ is the Bohr radius.
All error bars correspond to one standard deviation of the mean values.}
\label{fig2}
\end{center}
\end{figure}

\begin{figure*}
\begin{center}
\includegraphics[width=174mm]{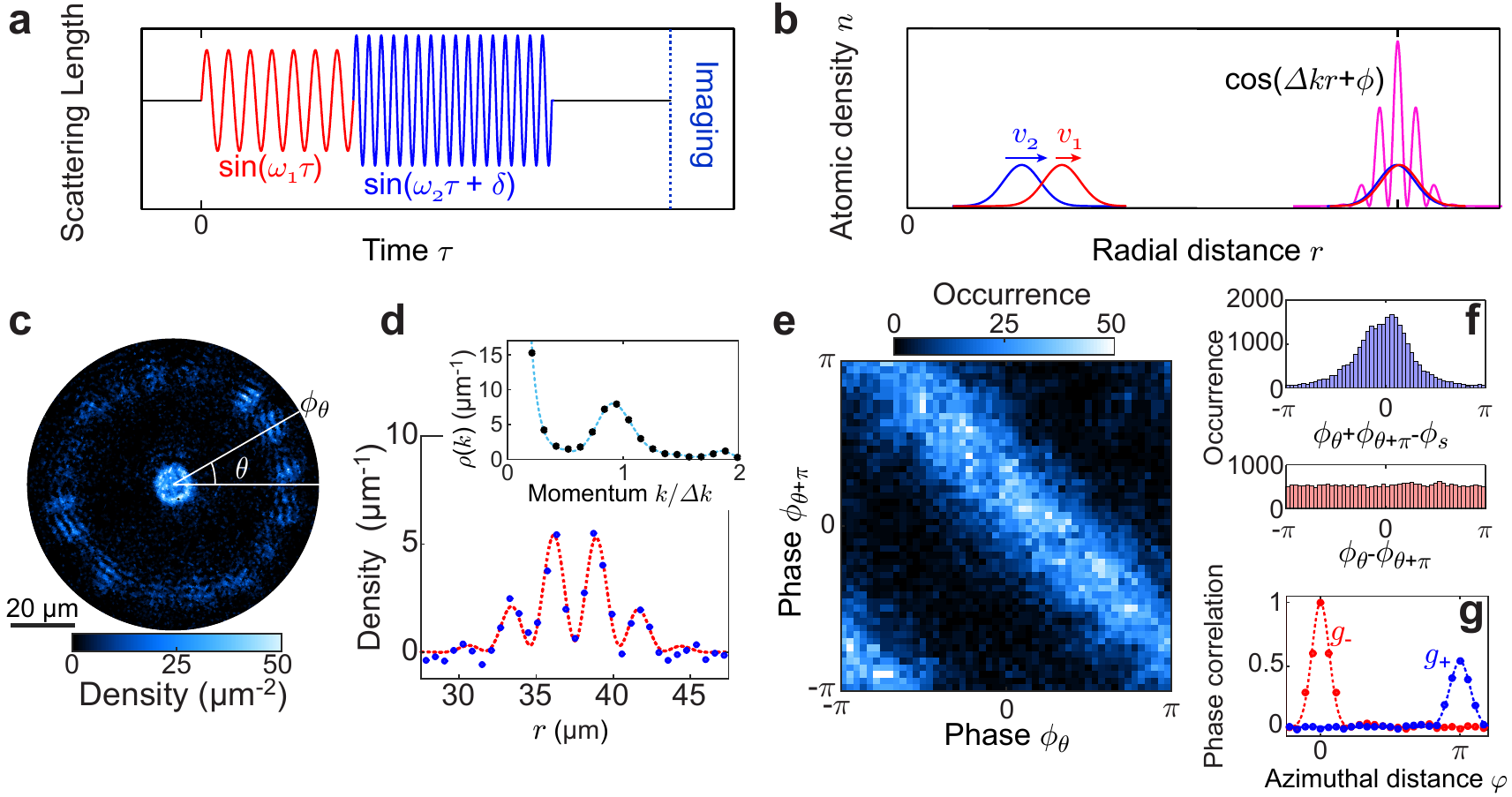}
\caption{\textbf{Long-range phase correlation of matterwave radiation.} Here the condensates are confined in a disk-shaped trap with radius $7~\mu$m. \textbf{a} illustrates the application of two pulses of scattering length modulation with frequencies  $\omega_1/2\pi$~=~3 and $\omega_2/2\pi$~=~5.63~kHz, and modulation amplitudes $a_{ac}$~=~56 and 72$a_0$. The relative phase of the pulses is $\delta$. \textbf{b}. The matter-wave jet created by the latter pulse propagates at a greater speed $v_2>v_1$ and interferes with atoms from the first pulse when they overlap. Here the matter-wave speeds are $v_i =\sqrt{\hbar\omega_i/m}$ for the $i-$th pulse. The interference is characterized by the wavenumber difference $\Delta k=k_2-k_1$, and the phase $\phi$. \textbf{c} shows an example interference pattern of the two radiation fields. The phase of the interference fringes $\phi_\theta$ is recorded as a function of the emission angle $\theta$. \textbf{d} shows the radial cut of the interference pattern, from which we determine the phase of the fringes based on Fourier transformation (See Methods). Dotted lines show guides to the eye. \textbf{e} and \textbf{f} show the concurrence of the extracted phases in the opposite directions, $\phi_\theta$ and $\phi_{\theta+\pi}$ for all emission angle $\theta$ from a collection of 200 images. A strong correlation of the two phases is described by $\phi_\theta+\phi_{\theta+\pi} = \phi_s$, where $\phi_s$~=~0.79(3) is obtained from fitting the data; $\phi_\theta-\phi_{\theta+\pi}$ appears to be random. \textbf{g} shows phase correlations $g_+$ (blue) and $g_-$ (red) between fringes separated by an angular distance $\varphi$, see Eq.~(6). Dots represent experimental data while dashed curves are guides to the eye (see Methods).
}
\label{fig3}
\end{center}
\end{figure*}

Our experiment starts with a Bose-Einstein condensate of $6\times 10^4$ atoms confined in a disk-shaped trap.
By modulating the magnetic field at frequency $\omega$ near a Feshbach resonance \cite{Chin2010,Tsatsos2017}, a jet-like two-dimensional emission of atoms with momentum $k_f=\sqrt{m\omega/\hbar}$ is observed few milliseconds after the modulation, where $m$ is the atomic mass. Such emission forms a fluctuating bosonic field, also called ``Bose fireworks'', and is a result of bosonic stimulation \cite{Logan2017,Feng2018}. Its evolution can be approximately described by the Hamiltonian in Eq.~(3) (see Methods).

In typical experiments, the emission carries as many as $276$ angular modes and each mode acquires a width of 1.33$^\circ$ (see Methods).
To study the distribution of mode population, we divide the emission pattern evenly into 180 angular slices.
For each slice, we extract the atom number $n$ and evaluate the probability distribution of the mode population $P(n)$ (see Fig.~\ref{fig2}\textbf{a}).

The measured mode population distributions well resemble that from a thermal radiation (see Fig.~\ref{fig2}\textbf{a}).
We extract the effective temperature $T$ based on a thermal model (see Methods), which fits the data excellently. Furthermore, the extracted temperature shows a clear linear dependence on the mean atomic population per mode $\bar n=(1/\xi)\int nP(n)dn$ with $\xi~=2^\circ/1.33^\circ~=~1.5$ the average number of modes within a $2^\circ$ slice (See Fig.~\ref{fig2}\textbf{b}).

The thermal distribution of the mode population can be understood in terms of Unruh effect.
The matter-wave field measured in our system simulates the vacuum state observed in an accelerating frame. We evaluate the simulated acceleration $A$ using Eq.~(\ref{eq4}) (Fig.~\ref{fig2}\textbf{b}), from which we can relate the temperature and the acceleration as $T~=~\kappa A/c$. From fitting the data, we obtain the ratio $\kappa~=~1.17(7)$ pK$\cdot$s. Our result agrees well with the Unruh prediction $\kappa=\hbar/2\pi k_\mathrm{B}\approx1.22$~pK$\cdot$s (see Eq.~(\ref{eq1})).

In addition to the temperature, we further evaluate the entropy per momentum mode $S=-k_{\mathrm{B}}\sum_{n} P(n)\ln P(n)+S_0$ (see Methods). Theoretically, the entropy should be the von Neumann entropy of the momentum mode after tracing out all others. Entropy is an important parameter to characterize black-hole thermodynamics \cite{Bekenstein1973,Hawking1976}.

For short modulation time $\tau<$~3~ms, the measured entropy is dominated by the detection noise $ S_\mathrm{b}=3.8~k_\mathrm{B}$. For long modulation duration $\tau$, the measured $S$ faithfully reflects the entropy of the matter-wave radiation. The entropy increases logarithmically with $A$ (see Fig.~\ref{fig2}\textbf{b}), consistent with the theory (see Methods).

\begin{figure*}
\begin{center}
\includegraphics[width=174mm]{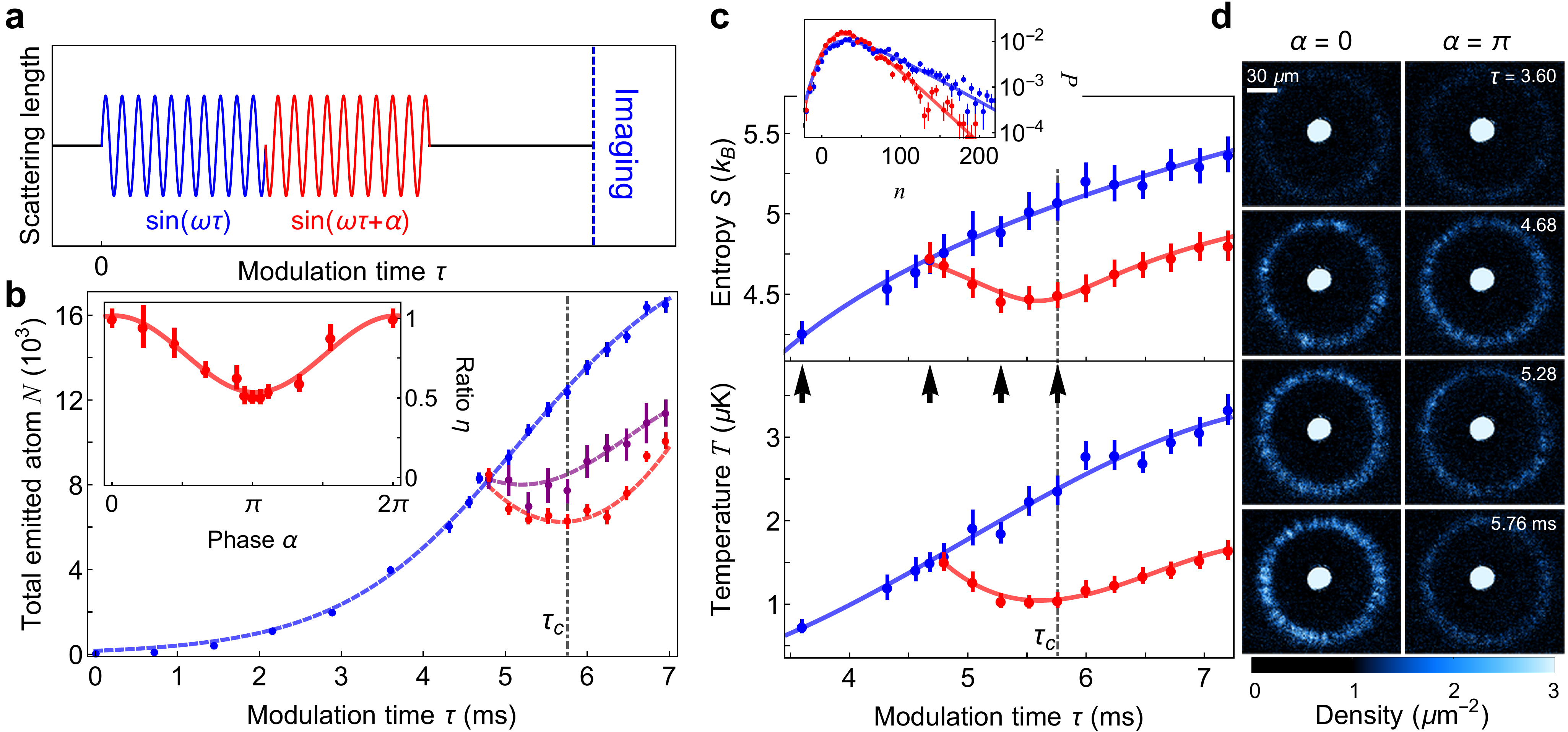}
\caption{\textbf{Time reversal of the matter-wave radiation field.} \textbf{a} shows that the scattering length is modulated at frequency $\omega/2\pi=2.1$~kHz with amplitude $a_{ac}=50$~$a_0$ for 4.75~ms before a phase jump $\alpha$ is introduced to the modulation. \textbf{b} shows the total emitted atom number $N$ versus total modulation time $\tau$. The blue, purple and red data correspond to the phase jumps of  $\alpha=0$, $2\pi/3$ and $\pi$, respectively. The dashed lines are guides to the eye. The inset shows the suppression ratio $\eta$ versus $\alpha$ evaluated at $\tau=\tau_c$. A sinusoidal fit gives the maximum reversal at $\alpha=0.98(3)\pi$ , where $\eta$ reaches 51(3)$\%$. \textbf{c} shows the entropy $S$ and temperature $T$ without $(\alpha=0$, blue circles$)$ and with the phase jump $(\alpha=\pi$, red circles$)$. The lines here are guides to the eye. The inset compares the population distributions $P(n)$ at $\tau_c$ with $\alpha=0$ (blue) and $\pi$ (red). The solid lines are the fits from our thermal model. \textbf{d} shows the average of 15 images of the matter-wave radiation at different times (indicated by arrows in panel \textbf{c}) with phase jump $\alpha=0$ or $\pi$. Here the condensates are confined in a disk-shaped trap with radius $13~\mu$m. All error bars correspond to one standard deviation of the mean value.}
\label{fig4}
\end{center}
\end{figure*}

While local measurements in our system seem to reveal a thermal distribution,
however, unlike incoherent black-body radiation, Hawking-Unruh radiation should exhibit both spatial and temporal coherence, reflecting its quantum origin.
In the following we investigate the coherence properties of the matter-wave radiation.

We first show the spatial coherence of the matter-wave field by probing the phase correlation between jets. For this, we perform a matter-wave interference experiment by applying two independent pulses of modulation on the scattering length; the first pulse has a lower frequency compared to the second one (see Fig.~\ref{3}\textbf{a}). The two frequencies are incommensurate to avoid influence from high-harmonic generations \cite{Feng2018}.
The pulses are arranged such that the atoms created by the second pulse leave the condensate later, but with a greater velocity than the atoms from the first pulse. When the two emitted waves overlap, they interfere and produce fringes (see Fig.~\ref{fig3}\textbf{b}). The phase of the fringes $\phi$ is given by the relative phase of the interfering matterwaves, and is dependent on the emission angle  $\theta$ (see Fig.~\ref{fig3}\textbf{c} and \textbf{d}).

We observe the phase correlation of fringes along counter-propagating directions. In Fig.~\ref{fig3}\textbf{e}, we present the occurrence distribution of the fringe phases in opposite directions, namely, $\phi_\theta$ and $\phi_{\theta+\pi}$.
The two phases correlate as $\phi_\theta +\phi_{\theta+\pi} = \phi_s$ (see Fig.~\ref{fig3}\textbf{e, f}) with $\phi_s$~=~0.79(3) consistent with our expectation(see Methods).

To be more quantitative, we evaluate the phase correlation function $g_\pm(\varphi)$ for all angular span $\varphi$ \cite{Langen2013}, defined as  (see Fig.~\ref{fig3}\textbf{g})
\begin{equation}
g_{\pm}(\varphi ) =| \langle e^{i\phi_{\theta}
\pm i\phi_{\theta+\varphi}}\rangle|.
\label{eq:g1}
\end{equation}
Here the angle brackets correspond to angular averaging over $\theta$ and ensemble averaging. The peak of $g_+$ at $\varphi = \pi$ confirms that fringe phases are only anti-correlated in the opposite directions. The lone peak of $g_-$ at $\varphi = 0$ shows the phase coherence within a single jet.

Since jets with different energy are generated independently, the correlations of the fringes indicate the phase correlations of counter-propagating jets with the same momentum.
Such phase correlation results from the coherent generation of atom pairs which are phase locked to the modulation; the correlation 
is also expected for the Unruh radiation \cite{Wald1994}, and resembles the phase coherence in the parametric down-conversion process in quantum optics \cite{Howell2004}.

Next we show the temporal coherence of the matter-waves radiation by reversing the time evolution.
Similar experiments to reverse parametric amplification are realized based on photonic and atomic fields with two well-defined outgoing modes and low atom numbers \cite{Zheng2013,Linnemann2016,Linnemann2017}, whereas the condensate in our system simultaneously couples to about 300 momentum modes, and involves about $10^5$ atoms.

Here we perform the experiment as follows: after modulating the scattering length, we jump the phase of the modulation by $\alpha$ (see Fig~\ref{fig4}\textbf{a}).
We monitor the evolution of the radiation patterns, from which we determine the total emitted atom number $N$ (Fig.~\ref{fig4}\textbf{b}). A clear suppression of atom number is shown for large phase jumps.
We evaluate the suppression ratio $\eta(\alpha)={N_\alpha (\tau_c)/ N_0 (\tau_c)}$ at time $\tau_c=5.76$~ms when the maximal reversal occurs (see Fig.~\ref{fig4}\textbf{b}).
In particular when $\alpha$ equals to $\pi$, the total excited atom number $N$ reduces by as much as $51(3)\%$ of that without the phase jump ($\alpha=0$).
At $\alpha=\pi$, a reversal of 26(3)$\%$ (or 2,200 atoms) of the matter-wave excitations back to the condensate is observed. Our results are consistent with the theoretical simulation (see Methods).

We evaluate the entropy $S$ and effective temperature $T$ from the distribution of emitted atom number, which remains thermal before and after the phase jump (Fig~4\textbf{c}). Here we compare them for the two cases with phase jump $\alpha=0$ and $\pi$. In the former case, $S$ and $T$ continuously increase while in the latter case, both of them decrease first but eventually increase again.
The reversal can be clearly seen from the strength of the emission pattern in the averaged images (see Fig.~\ref{fig4}\textbf{d}).
The reversal of these quantities suggests that the radiation originates from a unitary evolution.
The limited amount of reversal we can achieve is due to off-resonant coupling to the finite momentum modes close to $|\vec{k}|~=~k_f$ (see Methods).

In conclusion, we demonstrate a new type of quantum simulation to investigate quantum phenomena in a non-inertial frame.
By simulating vacuum in an accelerating frame, we observe the appearance of thermal radiation of matterwaves which resembles the Hawking-Unruh radiation. Such matter-wave radiation, albeit thermal from local measurements \cite{Kaufman2016}, possesses long-range spatial and temporal coherence, which distinguish it from classical thermal radiation. Quantum simulation of frame transformation can pave an alternate way to study the intriguing topics at the interface of quantum and relativistic physics \cite{Garay2000,Schutzhold2005,Horstmann2010,Bekenstein2015,Steinhauer2016} such as the quantization of field in a curved spacetime.

\noindent\textbf{Acknowledgement}

We thank R.~M. Wald, N.~D. Gemelke and L.W. Clark for helpful discussions and reading the manuscript. We thank K. Levin's group for providing the numerical server. We thank F. Fung for graphics preparation.
L. F. acknowledges support from a MRSEC-funded Graduate Research Fellowship. This work was partially supported by the University of Chicago Materials Research Science and Engineering Center, which is funded by the National Science Foundation under award number DMR-1420709, NSF grant PHY-1511696, and the Army Research Office-Multidisciplinary Research Initiative under grant W911NF-14-1-0003.

\bibliography{projectG}

\begin{thebibliography}{30}%
\makeatletter
\providecommand \@ifxundefined [1]{%
 \@ifx{#1\undefined}
}%
\providecommand \@ifnum [1]{%
 \ifnum #1\expandafter \@firstoftwo
 \else \expandafter \@secondoftwo
 \fi
}%
\providecommand \@ifx [1]{%
 \ifx #1\expandafter \@firstoftwo
 \else \expandafter \@secondoftwo
 \fi
}%
\providecommand \natexlab [1]{#1}%
\providecommand \enquote  [1]{``#1''}%
\providecommand \bibnamefont  [1]{#1}%
\providecommand \bibfnamefont [1]{#1}%
\providecommand \citenamefont [1]{#1}%
\providecommand \href@noop [0]{\@secondoftwo}%
\providecommand \href [0]{\begingroup \@sanitize@url \@href}%
\providecommand \@href[1]{\@@startlink{#1}\@@href}%
\providecommand \@@href[1]{\endgroup#1\@@endlink}%
\providecommand \@sanitize@url [0]{\catcode `\\12\catcode `\$12\catcode
  `\&12\catcode `\#12\catcode `\^12\catcode `\_12\catcode `\%12\relax}%
\providecommand \@@startlink[1]{}%
\providecommand \@@endlink[0]{}%
\providecommand \url  [0]{\begingroup\@sanitize@url \@url }%
\providecommand \@url [1]{\endgroup\@href {#1}{\urlprefix }}%
\providecommand \urlprefix  [0]{URL }%
\providecommand \Eprint [0]{\href }%
\providecommand \doibase [0]{http://dx.doi.org/}%
\providecommand \selectlanguage [0]{\@gobble}%
\providecommand \bibinfo  [0]{\@secondoftwo}%
\providecommand \bibfield  [0]{\@secondoftwo}%
\providecommand \translation [1]{[#1]}%
\providecommand \BibitemOpen [0]{}%
\providecommand \bibitemStop [0]{}%
\providecommand \bibitemNoStop [0]{.\EOS\space}%
\providecommand \EOS [0]{\spacefactor3000\relax}%
\providecommand \BibitemShut  [1]{\csname bibitem#1\endcsname}%
\let\auto@bib@innerbib\@empty
\bibitem [{\citenamefont {Hawking}(1974)}]{Hawking1974}%
  \BibitemOpen
  \bibfield  {author} {\bibinfo {author} {\bibfnamefont {S.~W.}\ \bibnamefont
  {Hawking}},\ }\bibfield  {title} {\enquote {\bibinfo {title} {Black hole
  explosions?}}\ }\href {http://www.nature.com/articles/248030a0} {\bibfield
  {journal} {\bibinfo  {journal} {Nature}\ }\textbf {\bibinfo {volume} {248}},\
  \bibinfo {pages} {30--31} (\bibinfo {year} {1974})}\BibitemShut {NoStop}%
\bibitem [{\citenamefont {Hawking}(1975)}]{Hawking1975}%
  \BibitemOpen
  \bibfield  {author} {\bibinfo {author} {\bibfnamefont {S.~W.}\ \bibnamefont
  {Hawking}},\ }\bibfield  {title} {\enquote {\bibinfo {title} {Particle
  creation by black holes},}\ }\href {\doibase 10.1007/BF02345020} {\bibfield
  {journal} {\bibinfo  {journal} {Communications in Mathematical Physics}\
  }\textbf {\bibinfo {volume} {43}},\ \bibinfo {pages} {199--220} (\bibinfo
  {year} {1975})}\BibitemShut {NoStop}%
\bibitem [{\citenamefont {Unruh}(1976)}]{Unruh1976}%
  \BibitemOpen
  \bibfield  {author} {\bibinfo {author} {\bibfnamefont {W.~G.}\ \bibnamefont
  {Unruh}},\ }\bibfield  {title} {\enquote {\bibinfo {title} {Notes on
  black-hole evaporation},}\ }\href {\doibase 10.1103/PhysRevD.14.870}
  {\bibfield  {journal} {\bibinfo  {journal} {Phys. Rev. D}\ }\textbf {\bibinfo
  {volume} {14}},\ \bibinfo {pages} {870--892} (\bibinfo {year}
  {1976})}\BibitemShut {NoStop}%
\bibitem [{\citenamefont {Wald}(1994)}]{Wald1994}%
  \BibitemOpen
  \bibfield  {author} {\bibinfo {author} {\bibfnamefont {Robert~M.}\
  \bibnamefont {Wald}},\ }\href@noop {} {\emph {\bibinfo {title} {Quantum Field
  Theory in Curved Spacetime and Black Hole Thermodynamics}}}\ (\bibinfo
  {publisher} {The University of Chicago Press},\ \bibinfo {year}
  {1994})\BibitemShut {NoStop}%
\bibitem [{\citenamefont {Bekenstein}(1973)}]{Bekenstein1973}%
  \BibitemOpen
  \bibfield  {author} {\bibinfo {author} {\bibfnamefont {Jacob~D.}\
  \bibnamefont {Bekenstein}},\ }\bibfield  {title} {\enquote {\bibinfo {title}
  {Black holes and entropy},}\ }\href {\doibase 10.1103/PhysRevD.7.2333}
  {\bibfield  {journal} {\bibinfo  {journal} {Phys. Rev. D}\ }\textbf {\bibinfo
  {volume} {7}},\ \bibinfo {pages} {2333--2346} (\bibinfo {year}
  {1973})}\BibitemShut {NoStop}%
\bibitem [{\citenamefont {Hawking}(1976{\natexlab{a}})}]{Hawking1976}%
  \BibitemOpen
  \bibfield  {author} {\bibinfo {author} {\bibfnamefont {S.~W.}\ \bibnamefont
  {Hawking}},\ }\bibfield  {title} {\enquote {\bibinfo {title} {Black holes and
  thermodynamics},}\ }\href {\doibase 10.1103/PhysRevD.13.191} {\bibfield
  {journal} {\bibinfo  {journal} {Phys. Rev. D}\ }\textbf {\bibinfo {volume}
  {13}},\ \bibinfo {pages} {191--197} (\bibinfo {year}
  {1976}{\natexlab{a}})}\BibitemShut {NoStop}%
\bibitem [{\citenamefont {Hayden}\ and\ \citenamefont
  {Preskill}(2007)}]{Hayden2007}%
  \BibitemOpen
  \bibfield  {author} {\bibinfo {author} {\bibfnamefont {Patrick}\ \bibnamefont
  {Hayden}}\ and\ \bibinfo {author} {\bibfnamefont {John}\ \bibnamefont
  {Preskill}},\ }\bibfield  {title} {\enquote {\bibinfo {title} {Black holes as
  mirrors: quantum information in random subsystems},}\ }\href
  {http://stacks.iop.org/1126-6708/2007/i=09/a=120} {\bibfield  {journal}
  {\bibinfo  {journal} {Journal of High Energy Physics}\ }\textbf {\bibinfo
  {volume} {2007}},\ \bibinfo {pages} {120} (\bibinfo {year}
  {2007})}\BibitemShut {NoStop}%
\bibitem [{\citenamefont {Giddings}(2013)}]{Giddings2013}%
  \BibitemOpen
  \bibfield  {author} {\bibinfo {author} {\bibfnamefont {Steven~B.}\
  \bibnamefont {Giddings}},\ }\bibfield  {title} {\enquote {\bibinfo {title}
  {Black holes, quantum information, and the foundations of physics},}\ }\href
  {\doibase 10.1063/PT.3.1946} {\bibfield  {journal} {\bibinfo  {journal}
  {Physics Today}\ }\textbf {\bibinfo {volume} {66}},\ \bibinfo {pages} {30}
  (\bibinfo {year} {2013})}\BibitemShut {NoStop}%
\bibitem [{\citenamefont {Clark}\ \emph {et~al.}(2017)\citenamefont {Clark},
  \citenamefont {Gaj}, \citenamefont {Feng},\ and\ \citenamefont
  {Chin}}]{Logan2017}%
  \BibitemOpen
  \bibfield  {author} {\bibinfo {author} {\bibfnamefont {Logan~W}\ \bibnamefont
  {Clark}}, \bibinfo {author} {\bibfnamefont {Anita}\ \bibnamefont {Gaj}},
  \bibinfo {author} {\bibfnamefont {Lei}\ \bibnamefont {Feng}}, \ and\ \bibinfo
  {author} {\bibfnamefont {Cheng}\ \bibnamefont {Chin}},\ }\bibfield  {title}
  {\enquote {\bibinfo {title} {Collective emission of matter-wave jets from
  driven bose-einstein condensates},}\ }\href
  {https://www.nature.com/articles/nature24272} {\bibfield  {journal} {\bibinfo
   {journal} {Nature}\ }\textbf {\bibinfo {volume} {551}},\ \bibinfo {pages}
  {356} (\bibinfo {year} {2017})}\BibitemShut {NoStop}%
\bibitem [{\citenamefont {Maldacena}(2018)}]{Maldacena1998}%
  \BibitemOpen
  \bibfield  {author} {\bibinfo {author} {\bibfnamefont {Juan}\ \bibnamefont
  {Maldacena}},\ }\bibfield  {title} {\enquote {\bibinfo {title} {The large n
  limit of superconformal field theories an supergravity},}\ }\href
  {http://dx.doi.org/10.4310/ATMP.1998.v2.n2.a1} {\bibfield  {journal}
  {\bibinfo  {journal} {Adv. Theor. Math. Phys.}\ ,\ \bibinfo {pages}
  {213--252}} (\bibinfo {year} {2018})}\BibitemShut {NoStop}%
\bibitem [{\citenamefont {Hawking}(1976{\natexlab{b}})}]{Hawking1976v2}%
  \BibitemOpen
  \bibfield  {author} {\bibinfo {author} {\bibfnamefont {S.~W.}\ \bibnamefont
  {Hawking}},\ }\bibfield  {title} {\enquote {\bibinfo {title} {Breakdown of
  predictability in gravitational collapse},}\ }\href {\doibase
  10.1103/PhysRevD.14.2460} {\bibfield  {journal} {\bibinfo  {journal} {Phys.
  Rev. D}\ }\textbf {\bibinfo {volume} {14}},\ \bibinfo {pages} {2460--2473}
  (\bibinfo {year} {1976}{\natexlab{b}})}\BibitemShut {NoStop}%
\bibitem [{\citenamefont {Susskind}(2006)}]{Susskind2006}%
  \BibitemOpen
  \bibfield  {author} {\bibinfo {author} {\bibfnamefont {Leonard}\ \bibnamefont
  {Susskind}},\ }\bibfield  {title} {\enquote {\bibinfo {title} {The paradox of
  quantum black holes},}\ }\href {http://www.nature.com/articles/nphys429}
  {\bibfield  {journal} {\bibinfo  {journal} {Nature Physics}\ }\textbf
  {\bibinfo {volume} {2}},\ \bibinfo {pages} {665--677} (\bibinfo {year}
  {2006})}\BibitemShut {NoStop}%
\bibitem [{\citenamefont {Almheiri}\ \emph {et~al.}(2013)\citenamefont
  {Almheiri}, \citenamefont {Marolf}, \citenamefont {Polchinski},\ and\
  \citenamefont {Sully}}]{Almheiri2013}%
  \BibitemOpen
  \bibfield  {author} {\bibinfo {author} {\bibfnamefont {Ahmed}\ \bibnamefont
  {Almheiri}}, \bibinfo {author} {\bibfnamefont {Donald}\ \bibnamefont
  {Marolf}}, \bibinfo {author} {\bibfnamefont {Joseph}\ \bibnamefont
  {Polchinski}}, \ and\ \bibinfo {author} {\bibfnamefont {James}\ \bibnamefont
  {Sully}},\ }\bibfield  {title} {\enquote {\bibinfo {title} {Black holes:
  complementarity or firewalls?}}\ }\href {\doibase 10.1007/JHEP02(2013)062}
  {\bibfield  {journal} {\bibinfo  {journal} {Journal of High Energy Physics}\
  }\textbf {\bibinfo {volume} {2013}},\ \bibinfo {pages} {62} (\bibinfo {year}
  {2013})}\BibitemShut {NoStop}%
\bibitem [{\citenamefont {Su}\ \emph {et~al.}(2017)\citenamefont {Su},
  \citenamefont {Ho}, \citenamefont {Mann},\ and\ \citenamefont
  {Ralph}}]{Su2016}%
  \BibitemOpen
  \bibfield  {author} {\bibinfo {author} {\bibfnamefont {Daiqin}\ \bibnamefont
  {Su}}, \bibinfo {author} {\bibfnamefont {C~T~Marco}\ \bibnamefont {Ho}},
  \bibinfo {author} {\bibfnamefont {Robert~B}\ \bibnamefont {Mann}}, \ and\
  \bibinfo {author} {\bibfnamefont {Timothy~C}\ \bibnamefont {Ralph}},\
  }\bibfield  {title} {\enquote {\bibinfo {title} {Quantum circuit model for
  non-inertial objects: a uniformly accelerated mirror},}\ }\href
  {http://stacks.iop.org/1367-2630/19/i=6/a=063017} {\bibfield  {journal}
  {\bibinfo  {journal} {New Journal of Physics}\ }\textbf {\bibinfo {volume}
  {19}},\ \bibinfo {pages} {063017} (\bibinfo {year} {2017})}\BibitemShut
  {NoStop}%
\bibitem [{\citenamefont {Chin}\ \emph {et~al.}(2010)\citenamefont {Chin},
  \citenamefont {Grimm}, \citenamefont {Julienne},\ and\ \citenamefont
  {Tiesinga}}]{Chin2010}%
  \BibitemOpen
  \bibfield  {author} {\bibinfo {author} {\bibfnamefont {Cheng}\ \bibnamefont
  {Chin}}, \bibinfo {author} {\bibfnamefont {Rudolf}\ \bibnamefont {Grimm}},
  \bibinfo {author} {\bibfnamefont {Paul}\ \bibnamefont {Julienne}}, \ and\
  \bibinfo {author} {\bibfnamefont {Eite}\ \bibnamefont {Tiesinga}},\
  }\bibfield  {title} {\enquote {\bibinfo {title} {Feshbach resonances in
  ultracold gases},}\ }\href {\doibase 10.1103/RevModPhys.82.1225} {\bibfield
  {journal} {\bibinfo  {journal} {Rev. Mod. Phys.}\ }\textbf {\bibinfo {volume}
  {82}},\ \bibinfo {pages} {1225--1286} (\bibinfo {year} {2010})}\BibitemShut
  {NoStop}%
\bibitem [{\citenamefont {Tsatsos}\ \emph {et~al.}(2017)\citenamefont
  {Tsatsos}, \citenamefont {Nguyen}, \citenamefont {Lode}, \citenamefont
  {Telles}, \citenamefont {Luo}, \citenamefont {Bagnato},\ and\ \citenamefont
  {Hulet}}]{Tsatsos2017}%
  \BibitemOpen
  \bibfield  {author} {\bibinfo {author} {\bibfnamefont {M.~C.}\ \bibnamefont
  {Tsatsos}}, \bibinfo {author} {\bibfnamefont {J.~H.~V.}\ \bibnamefont
  {Nguyen}}, \bibinfo {author} {\bibfnamefont {A.~U.~J.}\ \bibnamefont {Lode}},
  \bibinfo {author} {\bibfnamefont {G.~D.}\ \bibnamefont {Telles}}, \bibinfo
  {author} {\bibfnamefont {D.}~\bibnamefont {Luo}}, \bibinfo {author}
  {\bibfnamefont {V.~S.}\ \bibnamefont {Bagnato}}, \ and\ \bibinfo {author}
  {\bibfnamefont {R.~G.}\ \bibnamefont {Hulet}},\ }\bibfield  {title} {\enquote
  {\bibinfo {title} {Granulation in an atomic bose-einstein condensate},}\
  }\href@noop {} {\  (\bibinfo {year} {2017})},\ \Eprint
  {http://arxiv.org/abs/1707.04055} {arXiv:1707.04055} \BibitemShut {NoStop}%
\bibitem [{\citenamefont {Feng}\ \emph {et~al.}(2018)\citenamefont {Feng},
  \citenamefont {Hu}, \citenamefont {Clark},\ and\ \citenamefont
  {Chin}}]{Feng2018}%
  \BibitemOpen
  \bibfield  {author} {\bibinfo {author} {\bibfnamefont {Lei}\ \bibnamefont
  {Feng}}, \bibinfo {author} {\bibfnamefont {Jiazhong}\ \bibnamefont {Hu}},
  \bibinfo {author} {\bibfnamefont {Logan~W.}\ \bibnamefont {Clark}}, \ and\
  \bibinfo {author} {\bibfnamefont {Cheng}\ \bibnamefont {Chin}},\ }\bibfield
  {title} {\enquote {\bibinfo {title} {Complex correlations in high harmonic
  generation of matter-wave jets revealed by pattern recognition},}\
  }\href@noop {} {\  (\bibinfo {year} {2018})},\ \Eprint
  {http://arxiv.org/abs/1803.01786} {arXiv:1803.01786} \BibitemShut {NoStop}%
\bibitem [{\citenamefont {Langen}\ \emph {et~al.}(2013)\citenamefont {Langen},
  \citenamefont {Geiger}, \citenamefont {Kuhnert}, \citenamefont {Rauer},\ and\
  \citenamefont {Schmiedmayer}}]{Langen2013}%
  \BibitemOpen
  \bibfield  {author} {\bibinfo {author} {\bibfnamefont {T.}~\bibnamefont
  {Langen}}, \bibinfo {author} {\bibfnamefont {R.}~\bibnamefont {Geiger}},
  \bibinfo {author} {\bibfnamefont {M.}~\bibnamefont {Kuhnert}}, \bibinfo
  {author} {\bibfnamefont {B.}~\bibnamefont {Rauer}}, \ and\ \bibinfo {author}
  {\bibfnamefont {J.}~\bibnamefont {Schmiedmayer}},\ }\bibfield  {title}
  {\enquote {\bibinfo {title} {Local emergence of thermal correlations in an
  isolated quantum many-body system},}\ }\href
  {http://www.nature.com/articles/nphys2739} {\bibfield  {journal} {\bibinfo
  {journal} {Nature Physics}\ }\textbf {\bibinfo {volume} {9}},\ \bibinfo
  {pages} {640--643} (\bibinfo {year} {2013})}\BibitemShut {NoStop}%
\bibitem [{\citenamefont {Howell}\ \emph {et~al.}(2004)\citenamefont {Howell},
  \citenamefont {Bennink}, \citenamefont {Bentley},\ and\ \citenamefont
  {Boyd}}]{Howell2004}%
  \BibitemOpen
  \bibfield  {author} {\bibinfo {author} {\bibfnamefont {John~C.}\ \bibnamefont
  {Howell}}, \bibinfo {author} {\bibfnamefont {Ryan~S.}\ \bibnamefont
  {Bennink}}, \bibinfo {author} {\bibfnamefont {Sean~J.}\ \bibnamefont
  {Bentley}}, \ and\ \bibinfo {author} {\bibfnamefont {R.~W.}\ \bibnamefont
  {Boyd}},\ }\bibfield  {title} {\enquote {\bibinfo {title} {Realization of the
  einstein-podolsky-rosen paradox using momentum- and position-entangled
  photons from spontaneous parametric down conversion},}\ }\href {\doibase
  10.1103/PhysRevLett.92.210403} {\bibfield  {journal} {\bibinfo  {journal}
  {Phys. Rev. Lett.}\ }\textbf {\bibinfo {volume} {92}},\ \bibinfo {pages}
  {210403} (\bibinfo {year} {2004})}\BibitemShut {NoStop}%
\bibitem [{\citenamefont {Zheng}\ \emph {et~al.}(2013)\citenamefont {Zheng},
  \citenamefont {Ren}, \citenamefont {Wan},\ and\ \citenamefont
  {Chen}}]{Zheng2013}%
  \BibitemOpen
  \bibfield  {author} {\bibinfo {author} {\bibfnamefont {Yuanlin}\ \bibnamefont
  {Zheng}}, \bibinfo {author} {\bibfnamefont {Huaijin}\ \bibnamefont {Ren}},
  \bibinfo {author} {\bibfnamefont {Wenjie}\ \bibnamefont {Wan}}, \ and\
  \bibinfo {author} {\bibfnamefont {Xianfeng}\ \bibnamefont {Chen}},\
  }\bibfield  {title} {\enquote {\bibinfo {title} {Time-reversed wave mixing in
  nonlinear optics},}\ }\href {https://www.nature.com/articles/srep03245}
  {\bibfield  {journal} {\bibinfo  {journal} {Scientific Reports}\ }\textbf
  {\bibinfo {volume} {3}},\ \bibinfo {pages} {3245} (\bibinfo {year}
  {2013})}\BibitemShut {NoStop}%
\bibitem [{\citenamefont {Linnemann}\ \emph {et~al.}(2016)\citenamefont
  {Linnemann}, \citenamefont {Strobel}, \citenamefont {Muessel}, \citenamefont
  {Schulz}, \citenamefont {Lewis-Swan}, \citenamefont {Kheruntsyan},\ and\
  \citenamefont {Oberthaler}}]{Linnemann2016}%
  \BibitemOpen
  \bibfield  {author} {\bibinfo {author} {\bibfnamefont {D.}~\bibnamefont
  {Linnemann}}, \bibinfo {author} {\bibfnamefont {H.}~\bibnamefont {Strobel}},
  \bibinfo {author} {\bibfnamefont {W.}~\bibnamefont {Muessel}}, \bibinfo
  {author} {\bibfnamefont {J.}~\bibnamefont {Schulz}}, \bibinfo {author}
  {\bibfnamefont {R.~J.}\ \bibnamefont {Lewis-Swan}}, \bibinfo {author}
  {\bibfnamefont {K.~V.}\ \bibnamefont {Kheruntsyan}}, \ and\ \bibinfo {author}
  {\bibfnamefont {M.~K.}\ \bibnamefont {Oberthaler}},\ }\bibfield  {title}
  {\enquote {\bibinfo {title} {Quantum-enhanced sensing based on time reversal
  of nonlinear dynamics},}\ }\href {\doibase 10.1103/PhysRevLett.117.013001}
  {\bibfield  {journal} {\bibinfo  {journal} {Phys. Rev. Lett.}\ }\textbf
  {\bibinfo {volume} {117}},\ \bibinfo {pages} {013001} (\bibinfo {year}
  {2016})}\BibitemShut {NoStop}%
\bibitem [{\citenamefont {Linnemann}\ \emph {et~al.}(2017)\citenamefont
  {Linnemann}, \citenamefont {Schulz}, \citenamefont {Muessel}, \citenamefont
  {Kunkel}, \citenamefont {Prüfer}, \citenamefont {Frölian}, \citenamefont
  {Strobel},\ and\ \citenamefont {Oberthaler}}]{Linnemann2017}%
  \BibitemOpen
  \bibfield  {author} {\bibinfo {author} {\bibfnamefont {D}~\bibnamefont
  {Linnemann}}, \bibinfo {author} {\bibfnamefont {J}~\bibnamefont {Schulz}},
  \bibinfo {author} {\bibfnamefont {W}~\bibnamefont {Muessel}}, \bibinfo
  {author} {\bibfnamefont {P}~\bibnamefont {Kunkel}}, \bibinfo {author}
  {\bibfnamefont {M}~\bibnamefont {Prüfer}}, \bibinfo {author} {\bibfnamefont
  {A}~\bibnamefont {Frölian}}, \bibinfo {author} {\bibfnamefont
  {H}~\bibnamefont {Strobel}}, \ and\ \bibinfo {author} {\bibfnamefont {M~K}\
  \bibnamefont {Oberthaler}},\ }\bibfield  {title} {\enquote {\bibinfo {title}
  {Active su(1,1) atom interferometry},}\ }\href
  {http://stacks.iop.org/2058-9565/2/i=4/a=044009} {\bibfield  {journal}
  {\bibinfo  {journal} {Quantum Science and Technology}\ }\textbf {\bibinfo
  {volume} {2}},\ \bibinfo {pages} {044009} (\bibinfo {year}
  {2017})}\BibitemShut {NoStop}%
\bibitem [{\citenamefont {Kaufman}\ \emph {et~al.}(2016)\citenamefont
  {Kaufman}, \citenamefont {Tai}, \citenamefont {Lukin}, \citenamefont
  {Rispoli}, \citenamefont {Schittko}, \citenamefont {Preiss},\ and\
  \citenamefont {Greiner}}]{Kaufman2016}%
  \BibitemOpen
  \bibfield  {author} {\bibinfo {author} {\bibfnamefont {Adam~M.}\ \bibnamefont
  {Kaufman}}, \bibinfo {author} {\bibfnamefont {M.~Eric}\ \bibnamefont {Tai}},
  \bibinfo {author} {\bibfnamefont {Alexander}\ \bibnamefont {Lukin}}, \bibinfo
  {author} {\bibfnamefont {Matthew}\ \bibnamefont {Rispoli}}, \bibinfo {author}
  {\bibfnamefont {Robert}\ \bibnamefont {Schittko}}, \bibinfo {author}
  {\bibfnamefont {Philipp~M.}\ \bibnamefont {Preiss}}, \ and\ \bibinfo {author}
  {\bibfnamefont {Markus}\ \bibnamefont {Greiner}},\ }\bibfield  {title}
  {\enquote {\bibinfo {title} {Quantum thermalization through entanglement in
  an isolated many-body system},}\ }\href {\doibase 10.1126/science.aaf6725}
  {\bibfield  {journal} {\bibinfo  {journal} {Science}\ }\textbf {\bibinfo
  {volume} {353}},\ \bibinfo {pages} {794--800} (\bibinfo {year}
  {2016})}\BibitemShut {NoStop}%
\bibitem [{\citenamefont {Garay}\ \emph {et~al.}(2000)\citenamefont {Garay},
  \citenamefont {Anglin}, \citenamefont {Cirac},\ and\ \citenamefont
  {Zoller}}]{Garay2000}%
  \BibitemOpen
  \bibfield  {author} {\bibinfo {author} {\bibfnamefont {L.~J.}\ \bibnamefont
  {Garay}}, \bibinfo {author} {\bibfnamefont {J.~R.}\ \bibnamefont {Anglin}},
  \bibinfo {author} {\bibfnamefont {J.~I.}\ \bibnamefont {Cirac}}, \ and\
  \bibinfo {author} {\bibfnamefont {P.}~\bibnamefont {Zoller}},\ }\bibfield
  {title} {\enquote {\bibinfo {title} {Sonic analog of gravitational black
  holes in bose-einstein condensates},}\ }\href {\doibase
  10.1103/PhysRevLett.85.4643} {\bibfield  {journal} {\bibinfo  {journal}
  {Phys. Rev. Lett.}\ }\textbf {\bibinfo {volume} {85}},\ \bibinfo {pages}
  {4643--4647} (\bibinfo {year} {2000})}\BibitemShut {NoStop}%
\bibitem [{\citenamefont {Sch\"utzhold}\ and\ \citenamefont
  {Unruh}(2005)}]{Schutzhold2005}%
  \BibitemOpen
  \bibfield  {author} {\bibinfo {author} {\bibfnamefont {Ralf}\ \bibnamefont
  {Sch\"utzhold}}\ and\ \bibinfo {author} {\bibfnamefont {William~G.}\
  \bibnamefont {Unruh}},\ }\bibfield  {title} {\enquote {\bibinfo {title}
  {Hawking radiation in an electromagnetic waveguide?}}\ }\href {\doibase
  10.1103/PhysRevLett.95.031301} {\bibfield  {journal} {\bibinfo  {journal}
  {Phys. Rev. Lett.}\ }\textbf {\bibinfo {volume} {95}},\ \bibinfo {pages}
  {031301} (\bibinfo {year} {2005})}\BibitemShut {NoStop}%
\bibitem [{\citenamefont {Horstmann}\ \emph {et~al.}(2010)\citenamefont
  {Horstmann}, \citenamefont {Reznik}, \citenamefont {Fagnocchi},\ and\
  \citenamefont {Cirac}}]{Horstmann2010}%
  \BibitemOpen
  \bibfield  {author} {\bibinfo {author} {\bibfnamefont {B.}~\bibnamefont
  {Horstmann}}, \bibinfo {author} {\bibfnamefont {B.}~\bibnamefont {Reznik}},
  \bibinfo {author} {\bibfnamefont {S.}~\bibnamefont {Fagnocchi}}, \ and\
  \bibinfo {author} {\bibfnamefont {J.~I.}\ \bibnamefont {Cirac}},\ }\bibfield
  {title} {\enquote {\bibinfo {title} {Hawking radiation from an acoustic black
  hole on an ion ring},}\ }\href {\doibase 10.1103/PhysRevLett.104.250403}
  {\bibfield  {journal} {\bibinfo  {journal} {Phys. Rev. Lett.}\ }\textbf
  {\bibinfo {volume} {104}},\ \bibinfo {pages} {250403} (\bibinfo {year}
  {2010})}\BibitemShut {NoStop}%
\bibitem [{\citenamefont {Bekenstein}\ \emph {et~al.}(2015)\citenamefont
  {Bekenstein}, \citenamefont {Schley}, \citenamefont {Mutzafi}, \citenamefont
  {Rotschild},\ and\ \citenamefont {Segev}}]{Bekenstein2015}%
  \BibitemOpen
  \bibfield  {author} {\bibinfo {author} {\bibfnamefont {Rivka}\ \bibnamefont
  {Bekenstein}}, \bibinfo {author} {\bibfnamefont {Ran}\ \bibnamefont
  {Schley}}, \bibinfo {author} {\bibfnamefont {Maor}\ \bibnamefont {Mutzafi}},
  \bibinfo {author} {\bibfnamefont {Carmel}\ \bibnamefont {Rotschild}}, \ and\
  \bibinfo {author} {\bibfnamefont {Mordechai}\ \bibnamefont {Segev}},\
  }\bibfield  {title} {\enquote {\bibinfo {title} {Optical simulations of
  gravitational effects in the newton–schrödinger system},}\ }\href
  {https://www.nature.com/articles/nphys3451?cacheBust=1507919819525}
  {\bibfield  {journal} {\bibinfo  {journal} {Nature Physics}\ }\textbf
  {\bibinfo {volume} {11}},\ \bibinfo {pages} {872--878} (\bibinfo {year}
  {2015})}\BibitemShut {NoStop}%
\bibitem [{\citenamefont {Steinhauer}(2016)}]{Steinhauer2016}%
  \BibitemOpen
  \bibfield  {author} {\bibinfo {author} {\bibfnamefont {Jeff}\ \bibnamefont
  {Steinhauer}},\ }\bibfield  {title} {\enquote {\bibinfo {title} {Observation
  of quantum hawking radiation and its entanglement in an analogue black
  hole},}\ }\href {http://www.nature.com/articles/nphys3863} {\bibfield
  {journal} {\bibinfo  {journal} {Nature Physics}\ }\textbf {\bibinfo {volume}
  {12}},\ \bibinfo {pages} {959--965} (\bibinfo {year} {2016})}\BibitemShut
  {NoStop}%
\bibitem [{\citenamefont {Wang}\ \emph {et~al.}(2007)\citenamefont {Wang},
  \citenamefont {Hiroshima}, \citenamefont {Tomita},\ and\ \citenamefont
  {Hayashi}}]{wang2007}%
  \BibitemOpen
  \bibfield  {author} {\bibinfo {author} {\bibfnamefont {Xiang-Bin}\
  \bibnamefont {Wang}}, \bibinfo {author} {\bibfnamefont {Tohya}\ \bibnamefont
  {Hiroshima}}, \bibinfo {author} {\bibfnamefont {Akihisa}\ \bibnamefont
  {Tomita}}, \ and\ \bibinfo {author} {\bibfnamefont {Masahito}\ \bibnamefont
  {Hayashi}},\ }\bibfield  {title} {\enquote {\bibinfo {title} {Quantum
  information with gaussian states},}\ }\href {\doibase
  https://doi.org/10.1016/j.physrep.2007.04.005} {\bibfield  {journal}
  {\bibinfo  {journal} {Physics Reports}\ }\textbf {\bibinfo {volume} {448}},\
  \bibinfo {pages} {1 -- 111} (\bibinfo {year} {2007})}\BibitemShut {NoStop}%
\bibitem [{\citenamefont {Clark}\ \emph {et~al.}(2018)\citenamefont {Clark},
  \citenamefont {Anderson}, \citenamefont {Feng}, \citenamefont {Gaj},
  \citenamefont {Levin},\ and\ \citenamefont {Chin}}]{GaugeFieldPaper}%
  \BibitemOpen
  \bibfield  {author} {\bibinfo {author} {\bibfnamefont {Logan~W.}\
  \bibnamefont {Clark}}, \bibinfo {author} {\bibfnamefont {Brandon~M.}\
  \bibnamefont {Anderson}}, \bibinfo {author} {\bibfnamefont {Lei}\
  \bibnamefont {Feng}}, \bibinfo {author} {\bibfnamefont {Anita}\ \bibnamefont
  {Gaj}}, \bibinfo {author} {\bibfnamefont {Kathy}\ \bibnamefont {Levin}}, \
  and\ \bibinfo {author} {\bibfnamefont {Cheng}\ \bibnamefont {Chin}},\
  }\bibfield  {title} {\enquote {\bibinfo {title} {Observation of
  density-dependent gauge fields in a bose-einstein condensate based on
  micromotion control in a shaken two-dimensional lattice},}\ }\href@noop {} {\
   (\bibinfo {year} {2018})},\ \Eprint {http://arxiv.org/abs/1801.10077}
  {arXiv:1801.10077} \BibitemShut {NoStop}%
\end{thebibliography}%

\clearpage

\clearpage
\widetext
\setcounter{equation}{0}
\setcounter{figure}{0}
\setcounter{table}{0}
\setcounter{page}{1}
\makeatletter
\renewcommand{\theequation}{M\arabic{equation}}
\renewcommand{\thefigure}{M\arabic{figure}}
\renewcommand{\thetable}{M\arabic{table}}
\begin{Large}
\begin{center}
Methods for

\textbf{Quantum Simulation of Coherent Hawking-Unruh radiation}
\end{center}
\end{Large}

\section{Equivalence of the time evolution and Rindler frame transformation}
The time evolution under the Hamiltonian $\mathcal H=i\hbar\sum_k g_k(a^\dagger_k a^\dagger_{-k}-a_k a_{-k})$ can be solved analytically by equations of motion in the Heisenberg picture,
\begin{align}
\dot{a}_k &= \frac{i}{\hbar}[H,a_k] = g_k a^{\dagger}_{-k}\\
\dot{a}_{-k} &= \frac{i}{\hbar}[H,a_{-k}] = g_k a^{\dagger}_k.
\end{align}
Then we get the expressions for the time evolution of the operators as
\begin{align}
\begin{bmatrix}
a_k(\tau)\\
a_{-k}^{\dagger}(\tau)
\end{bmatrix}
= \begin{bmatrix}
\cosh(g_k\tau) &\sinh(g_k\tau)\\
\sinh(g_k\tau) &\cosh(g_k\tau)
\end{bmatrix}
\begin{bmatrix}
a_k(0)\\
a_{-k}^{\dagger}(0)
\end{bmatrix}.
\label{eqM3}
\end{align}
This result shares the same form of the Bogoliubov transformation which is related to the spontaneous particle creation from the vaccum, the Hawking-Unruh effects and the squeezed states of light.

On the other hand, if we look at the problem of boosting a quantized scalar field into an accelerating frame with acceleration $A$, it can be realized by the Rindler transformation \cite{Su2016} as
\begin{align}
\begin{bmatrix}
\hat{b}^{R}_{\omega}\\
\hat{b}^{\dagger L}_{\omega}
\end{bmatrix}
=
\begin{bmatrix}
\cosh(r_{\omega}) &\sinh(r_{\omega})\\
\sinh(r_{\omega}) &\cosh(r_{\omega})
\end{bmatrix}
\begin{bmatrix}
\hat{c}_{m\omega}\\
\hat{d}^{\dagger}_{m\omega}
\end{bmatrix},
\label{eqM4}
\end{align}
where $(\hat{b}^{R}_{\omega},\hat{b}^{L}_{\omega})$ are the annihilation operators in the accelerating frame, $L$ and $R$ correspond to two Rindler wedges propagating along two different directions, $(\hat{c}_{\omega}, \hat{d}_{\omega})$ are the annihilation operators of Unruh modes whose vacuum is the Minkovski vacuum in the inertial frame. The parameter $r_{\omega}$ satisfies the equation of $\tanh r_{\omega} = e^{-\pi\omega c/A}$.

We compare Eq. (\ref{eqM3}) with (\ref{eqM4}) and find the similarity in their expressions. The equivalence can be built by treating the operators at time $ \tau$ as the Rindler operators in a non-inertial frame, which leads to $g_k\tau = r_{\omega/2}$. Thus, we obtain
\begin{equation}
g_{k} = \frac{1}{2\tau}\ln\coth(\frac{\pi\omega c}{4A}),
\label{gk}
\end{equation}
corresponding the coupling strength of mode $k$ generating the Unruh radiation in the frequency component $\omega/2$.

\section{Hamiltonian and evolution of the condensate with a modulated interaction}
We start with the second quantization process of the Hamiltonian
\begin{equation}
H=\int d^3\bold{r}\Psi^\dagger(\bold r,t)\frac{p^2}{2m}\Psi(\bold r,t)+{\tilde g(t)\over 2}\int d^3\bold{r}\Psi^\dagger(\bold r,t)\Psi^\dagger(\bold r,t)\Psi(\bold r,t)\Psi(\bold r,t),
\end{equation}
where $\tilde g(t)=4\pi\hbar^2a(t)/m$ is the coupling strength and proportional to the scattering length $a(t)$.
In a driven condensate, the scattering length is in the form of $a(t)=a_{dc}+a_{ac}\sin(\omega t)$, where $\omega$ is the modulation frequency. By applying the Fourier transformation of the field operator
\begin{equation}
\Psi(\bold r,t)={1\over\sqrt V}\sum_k e^{i\bold k \bold r} a_{\bold k},
\end{equation}
where $V$ is the volume of the condensate, we obtain the Hamiltonian in the momentum space as
\begin{equation}
H=\sum_{\bold k}\epsilon_k a_\vk^\dagger a_\vk+ {\tilde g(t)\over 2V}\sum_{\bold k_1,\bold k_2,\Delta\bold k}a^\dagger_{\bold k_1+\Delta \bold k}a^\dagger_{\bold k_2-\Delta \bold k}a_{\bold k_1}a_{\bold k_2}.
\end{equation}

By entering the interaction picture and eliminating $H_0=\sum_{\bold k}\epsilon_k a_\vk^\dagger a_\vk$, we simplify the Hamiltonian under the rotating wave and Bogoliubov approximations and only keep the resonant terms. Then the interaction Hamiltionian becomes
\begin{equation}
    H = i\hbar g \sum_{|\vk|=k_f}(a_k^\dagger a_{-k}^\dagger-a_k a_{-k}),
\label{Two-mode}
\end{equation}
where $g = \pi\hbar N_0 a_{ac}/mV$ and $k_f = \sqrt{m\omega/\hbar}$.

According to Eq. \eqref{eqM3}, we know the evolution of the operators as
\begin{align}
\begin{bmatrix}
a_k(\tau)\\
a^{\dagger}_{-k}(\tau)
\end{bmatrix}
=
e^{g\tau\sigma_x}
\begin{bmatrix}
a_k(0)\\
a^{\dagger}_{-k}(0)
\end{bmatrix}.
\end{align}
where $\sigma_x$ is the x-component of Pauli matrices. In order to simulate the Rindler transformation, we have to match the form of Eq. (\ref{eqM4}) and obtain the simulated acceleration $A$ as
\begin{equation}
A = {\pi\omega c\over 2\ln\coth(g\tau)}.
\end{equation}

The mean population per mode increases as $\Bar{n} = \langle a^{\dagger}_k(\tau) a_k(\tau) \rangle = \sinh^2(g\tau) $. Thus, $A$ can be characterized by the mean population per mode $\bar n$ as
\begin{equation}
    A = \frac{2\pi c E_{k_f}}{\hbar \ln(1+1/\bar n)}\xrightarrow[\bar n \gg1]{} \frac{2\pi cE_{k_f}}{\hbar}\bar n,
    \label{eq13}
\end{equation}
where $E_{k_f} = \hbar\omega/2$ is the kinetic energy of each atom.

Now, let's consider the evolution of the wave function instead of the operators. Two counter-propagating modes with momentum $k$ and $-k$ are generated together and we have to consider them at the same time. By grouping $k$ and $-k$ together, we decompose the Hamiltonian into $H=\sum h_k$, where $h_k=i\hbar g(a_k^\dagger a_{-k}^\dagger-a_k a_{-k})$. Then we only need to consider the evolution of each $h_k$. To simplify the notation without loss of the generality, we use $h$ to replace $h_k$. Therefore, the evolution of the wave function can be written as \cite{wang2007}
\begin{equation}
    \ket{\psi(\tau)} = e^{-i h\tau/\hbar}\ket{0}
    =\frac{1}{\cosh(g\tau)}\sum_{n=0}^{\infty}\tanh^n(g\tau)\ket{n,n}.
\end{equation}
The wave function matches the Minkovski vacuum expanding in the basis of the Rindler frame \cite{Wald1994}. The density matrix of one single mode such as $k$ is determined by tracing out the other mode $-k$, \textit{i.e.}
\begin{equation}
    \hat\rho_k (\tau) = \mathrm{Tr}_{-k}{\ket{\psi(\tau)}_I\bra{\psi(\tau)}_I} = \sum_{n=0}^{\infty}p_n\ket{n}_k\bra{n}_k
\end{equation}
where $p_n= \tanh^{2n}(g\tau)/\cosh^2(g\tau)$. By comparing with a thermal distribution of an ideal Bose gas
\begin{equation}
\tilde p(T)=e^{-{n\hbar\omega\over 2k_\mathrm{B}T}}(1-e^{-{\hbar\omega\over 2k_\mathrm{B}T}}),
\end{equation}
we can build a direct mapping between the effective temperature with the time $\tau$ or the mean population $\bar n$ as
\begin{align}
T &= \frac{E_{k_f}}{2k_\mathrm{B}\ln\coth(g\tau) } \\
&= \frac{E_{k_f}}{k_\mathrm{B}\ln(1+1/\bar n)} \xrightarrow[\bar n \gg 1]{}\frac{E_{k_f}}{k_\mathrm{B}}\bar n.
\label{eq17}
\end{align}
where the mean population
\begin{align}
    \bar n = \sum_{n=0}^{\infty}np_n = \frac{1}{e^{E_{k_f}/k_{\mathrm{B}}T}-1}
\end{align}
follows the Bose-Einstein statistics.

We also characterize our system by the entropy $S$. The von Neumann entropy of the thermal distribution is
\begin{align}
S =& -k_{\mathrm{B}}\mathrm{Tr}\left(\hat\rho_k\ln\hat\rho_k\right)\nonumber\\
=&k_{\mathrm{B}}[\ln\Bar{n}+(\Bar{n}+1)\ln(1+1/\Bar{n})],
\label{eq16}
\end{align}
which is the solid purple line (theory without detection noise) plotted in Fig~2\textbf{b}.

The mean atom number as a function of acceleration A is $\bar n = 1/(e^{2\pi cE_{k_f}/\hbar A}-1)$. We insert it into Eq. \eqref{eq16} and get the relation between $S$ and $A$ as
\begin{equation}
    S = -k_\mathrm{B}\left[\ln\left(e^{2\pi cE_{k_f}/\hbar A}-1\right)-\frac{2\pi cE_{k_f}/\hbar A}{(1-e^{-2\pi cE_{k_f}/\hbar A})}\right].
\end{equation}

When $\Bar{n}\gg1$, the entropy is approximated as $S = k_{\mathrm{B}}\ln(e\Bar{n})$, where $e=2.718\ldots$. Using Eq. \eqref{eq13}, we obtain the entropy $S$ dependence on $A$ under large $\bar n$ limit as
\begin{equation}
S = k_{\mathrm{B}}\ln\left(\frac{e\hbar A}{2\pi cE_{k_f}}\right),
\end{equation}
in which $S$ increases logarithmically with $A$.\\

\section{Determination of mode width and effective temperature}

In this section, we first determine the mode width experimentally. In Ref. \cite{Logan2017}, the measurement of second-order correlation function $g^{(2)}(\theta)$ of Bose fireworks was reported. We have $g^{(2)}(0)=2$ indicating that in one mode there is a relation of $\Delta n^2_M=[g^{(2)}(0)-1]\langle n_M\rangle^2$ where $\langle n_M\rangle$ and $\Delta n^2_M$ are the mean and variance of the atom number.  

\begin{figure}[h!]
\begin{center}
\includegraphics[width=182mm]{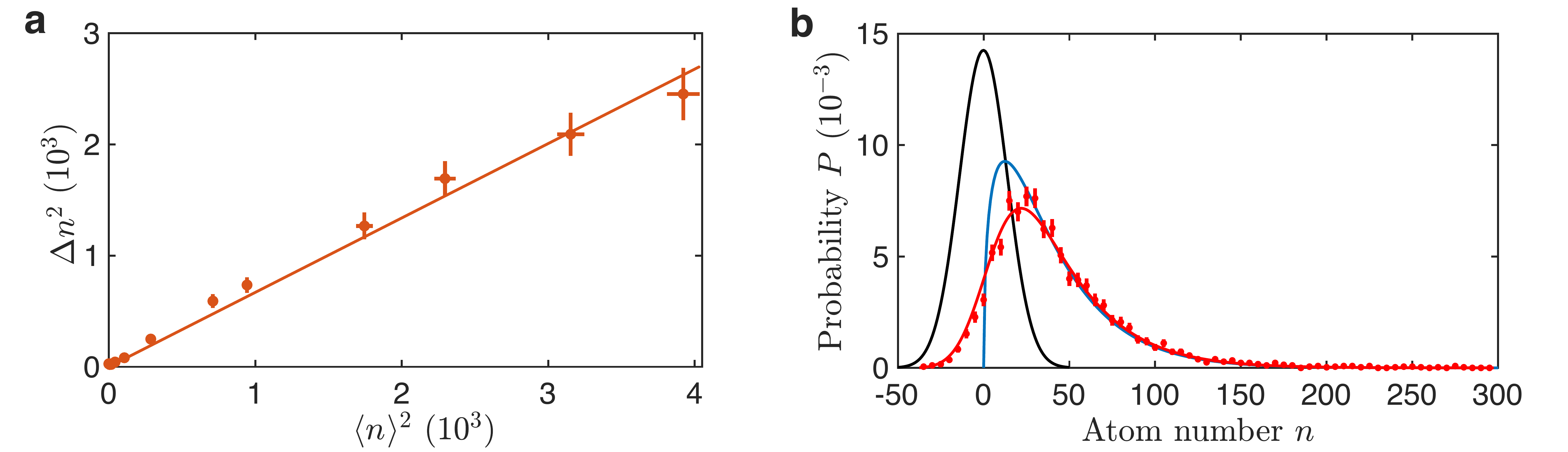}
\caption{\textbf{Determination of the mode width and the fitting of the measured probability distribution P(n)}. \textbf{a} shows the linear dependence of mean square $\mean{n}^2$ and variance $\Delta n^2$ of atom number distribution in the slice with an angular width $\Delta\theta_S = 2^{\circ}$, from which we have subtracted the contribution from the detection noise. \textbf{b} shows the background atom number distribution $G(n,1.5)$ (black line), ideal emitted atom number distribution $p(n,1.5)$ (blue line) and the convolution between both of them $P(n,1.5)$ (red line) which fits the measured probability distribution $P(n)$ (red circles) at the modulation time $\tau = 4.8$~ms.   }
\label{S1}
\end{center}
\end{figure}

Experimentally we slice our emission patterns into 180 slices and count the atom number in each slice. Based on the histogram of atom counting from the measurements, we build the probability distribution function $P(n)$ and calculate $\langle n\rangle=\int nP(n)dn$ and $\Delta n^2=\langle n^2\rangle-\langle n\rangle^2-\Delta n^2_{noise}$. Here $\Delta n^2_{noise}$ is the variance contributed from the detection noise which is statistically independent with the signal from atom counting. From that, we find a linear dependence between the mean atom number square and the variance from the experiment as (see Fig.~\ref{S1})
\begin{equation}
    \langle n\rangle^2 = \xi \Delta n^2.
\end{equation}
Here $\xi=\Delta\theta_S/\Delta\theta_J=1.49(7)$ is determined from the fitting, characterizing the mode width $\Delta \theta_J$ with each slice's width $\Delta\theta_S=2^\circ$. Therefore, $\Delta\theta_J$ equals to $1.33^\circ$. We also calculate $\Delta\theta_J$ from another independent way. Using the formula $\Delta\theta_J=1.62/(Rk_f)$ in Ref.~\cite{Logan2017} which comes from the half width at half maximum of the peak at $\phi = 0$ in the $g^{(2)}$ function, we obtain a consistent result of $\Delta\theta_J=1.3^\circ$.

To test and verify
that the emitted atom number in each mode follows a thermal distribution, we derive a more general formula for the probability distribution $p(n,\xi)$ in a slice with any width $\Delta\theta_S=\xi\Delta\theta_J$.
Because the mean population per mode $\langle n_M\rangle$ is always larger than 1 in our measurements, we treat the distribution $p(n,\xi)$ as a continuous function where the summation $\sum_{n=0}^{\infty} p(n,\xi)=1$ is replaced by an integral $\int_{0}^{\infty} dn p(n,\xi)=1$.

Here we would like to list a few properties of the function $p(n,\xi)$. First, $p(n,\xi)$ must equal to 0 when $n$ is a negative number. Second, if the angular slice only contains one momentum mode (\textit{i.e.} $\xi = 1$), $p(n,1)$ should be a thermal distribution, where $p(n,1)$ equals $\beta e^{-\beta n}$ with $\beta = E_{k_f}/k_{\mathrm{B}}T$.  Third, $p(n,\xi)$ have to satisfy the addition rule that combining two slices of $\xi_1$ and $\xi_2$ will create a new slice of $\xi_1+\xi_2$. We can write the third requirement more explicitly as a mathematical equation
\begin{equation}
p(n,\xi_1+\xi_2) = \int_{-\infty}^\infty p(n^\prime,\xi_1) p(n-n^\prime,\xi_2)dn^\prime.
\end{equation}
From all the above conditions, we solve the probability distribution $p(n,\xi)$ analytically as
\begin{eqnarray}
p(n,\xi ) &=&
\begin{cases}
\beta^\xi n^{\xi-1}e^{-\beta n}/\Gamma(\xi) & n\geq 0\\
0 & n<0,
\end{cases}
\end{eqnarray}\label{eqn:prob_slice}
where $\Gamma(\xi)$ is the gamma function.

In addition to the signals from the atoms, the detection noise contributes to the measured probability distribution of the atom number. Experimentally we characterize this noise distribution $G(n,\xi)$ by inspecting the images without any radiations. Once we get $G(n,\xi)$, we convolve it with $p(n,\xi)$ to get a full distribution function
\begin{equation}
    P(n,\xi)=\int_{-\infty}^\infty d n^\prime p(n^\prime,\xi)G(n-n^\prime,\xi).
\end{equation}
Then we use this function to fit our data extracting out the temperature $T$ under the condition of $\xi=1.5$ (see Fig.~\ref{S1}\textbf{b}).

\section{Characterization of entropy from population distribution}

We define the entropy in one slice with the width of $\xi\Delta\theta_J$ as $S(\xi)$. First we use the probability distribution $p(n,\xi)$ derived in the previous section to evaluate $S(\xi)$, which gives
\begin{equation}
    S(\xi)/k_\mathrm{B}=-\int_{-\infty}^\infty dn p(n,\xi)\ln p(n,\xi)=-\ln\beta+\xi+\ln\Gamma(\xi)-(\xi-1)\Gamma^\prime(\xi)/\Gamma(\xi).\label{eq26}
\end{equation}

In our data analysis, we divide the radiation pattern into 180 slices and determine the probability distribution $P(n)$. Thus, the entropy directly measured by the experiment is
\begin{equation}
    S(1.5)=-k_\mathrm{B}\sum_n P(n)\ln P(n).
\end{equation}
We show that the entropy in a single mode $S(1)$ is given by
\begin{equation}
    S(1)=S(1.5)-S_0=-k_\mathrm{B}\sum_n P(n)\ln P(n)-S_0
\end{equation}
based on Eq. (\ref{eq26}) where $S_0=k_{\mathrm{B}}\left[\xi-1+\ln\Gamma(\xi)-(\xi-1)\Gamma^\prime(\xi)/\Gamma(\xi)\right]\vert_{\xi=1.5}=0.37k_\mathrm{B}$.

For the theoretical curve with noise plotted in Fig.~\ref{fig2}\textbf{b} (blue solid line), we characterize the detection noise per mode $ G(n,1)$ and then evaluate the theoretical distribution by convolving $ G (n,1)$ with $p(n,1)$ as
\begin{equation}
    \tilde P(n,1)=\int_{-\infty}^\infty dn^\prime p(n^\prime,1) G(n-n^\prime,1).
\end{equation}
And we calculate the entropy as
\begin{equation}
    S=-k_{\mathrm{B}}\int dn \tilde P(n,1)\ln \tilde P(n,1),
\end{equation}
which matches our experimental data (see the blue solid line in Fig.~\ref{fig2}\textbf{b}).

\section{Phase correlations of atomic radiation field}

Here we calculate the phase correlations between interference fringes, which directly relate to that between emitted jets. We consider two sets of independent jets which are generated by two pulses of scattering length modulation with certain phase. In the interaction picture, the wave function can be written as $\ket{\psi}_I=\ket{\psi^{(1)}}_I\otimes\ket{\psi^{(2)}}_I$. Each $\ket{\psi^{(j)}}_I$ follows
\begin{equation}
    \ket{\psi^{(j)}}_I={1\over\cosh(\gamma_j)}\sum_{n=0}^\infty\left[e^{i(\phi_{M_j}-\pi/2)}\tanh(\gamma_j)\right]^n |n,n\rangle_{k_j,-k_j}
\end{equation}
under the Hamiltonian
\begin{equation}
    H^{(i)}_I=g_j e^{i\phi_{M_j}} a^\dagger_{k_j} a^\dagger_{-k_j} +g_j e^{-i\phi_{M_j}} a_{k_j} a_{-k_j}
\end{equation}
where $\phi_{M_j}$ is given by the phase of external driving field, $\gamma_j=g_j\tau_j$ and $\tau_i$ is the modulation duration of the pulse.

To take the dynamical phase into account, we convert the wave function back to Schr\"odinger's picture, and the wave function is written as
\begin{eqnarray}
|\psi\rangle_S &=& |\psi^{(1)}\rangle_S \otimes |\psi^{(2)}\rangle_S ,
\end{eqnarray}
where $|\psi^{(j)}\rangle_S$ is given by
\begin{eqnarray}
|\psi^{(j)}\rangle_S &=&e^{-iH_0^{(j)}t/\hbar}|\psi^{(j)}\rangle_I\nonumber\\
&=&\frac{1}{\cosh(\gamma_j)}\sum^{\infty}_{n=0}\left[e^{i(\phi_{M_j}-\omega_jt-\pi/2)}\tanh(\gamma_j)\right]^n |n,n\rangle_{k_j,-k_j}.
\end{eqnarray}
Here $H_0^{(i)} = \hbar\omega_i(a_{\vk_j}^\dagger a_{\vk_j}+a_{-\vk_j}^\dagger a_{-\vk_j})/2$ is energy term which was previously eliminated in the interaction picture.

The interference operators between the two sets of jets are $\hat{I}_f = a_{\vk_1}a_{\vk_2}^\dagger$ and $\hat{I}_b = a_{-\vk_1}a_{-\vk_2}^\dagger$ which correspond to the forward and backward directions. We introduce four more interference operators
as $\hat{I}_{j+}=a_{\vk_j}a_{-\vk_j}$ and $\hat{I}_{j-}=a_{\vk_j}a^\dagger_{-\vk_j}$ with $j$~=~1 or 2. The mean value for the interference operator $\hat{I}_{i\pm}$ is evaluated as
\begin{eqnarray}
\langle \hat{I}_{j+}\rangle &=&\langle\psi^{(j)}|_S \left(a_{\vk_j} a_{-\vk_j}\right) |\psi^{(j)}\rangle_S\nonumber\\
&=& \sqrt{\langle n_j\rangle(\langle n_j\rangle+1)}e^{i(\phi_{M_j}-\omega_jt-\pi/2)}\label{eq:Iplus}\\
\langle \hat{I}_{j-}\rangle &=&\langle\psi^{(j)}|_S \left(a_{\vk_j} a_{-\vk_j}^\dagger\right) |\psi^{(j)}\rangle_S\nonumber\\
&=& 0,\label{eq:Iminus}
\end{eqnarray}
where $\langle n_j\rangle$ is the mean atom number in each set of jets.

Phase correlation between interference fringes can be directly decomposed into the interference operators in each set of jets. The phase correlation $g_+(\theta = \pi)$ is proportional to the correlation between $\hat{I}_f$ and $\hat{I}_b$ , together with Eq.~(\ref{eq:Iplus}) we get
\begin{eqnarray}
\langle e^{i(\phi_{\theta}+\phi_{\theta+\pi})}\rangle \propto \langle \hat{I}_f\hat{I}_b \rangle
&=& \langle\psi^{(1)}|_S \otimes \langle\psi^{(2)}|_S \left(a_{\vk_1} a^\dagger_{\vk_2} a_{-\vk_1} a_{-\vk_2}^\dagger\right)  |\psi^{(1)}\rangle_S \otimes |\psi^{(2)}\rangle_S \nonumber\\
&=&\langle\psi^{(1)}|_S \left(a_{\vk_1} a_{-\vk_1}\right) |\psi^{(1)}\rangle_S \langle\psi|_S^{(2)}\left( a_{\vk_2}^\dagger a_{-\vk_2}^\dagger\right) |\psi\rangle_S^{(2)}\nonumber\\
&=& \langle\hat{I}_{1+}\rangle\langle \hat{I}_{2+}^\dagger\rangle\nonumber\\
&=&\sqrt{\langle n_1\rangle(\langle n_1\rangle+1)}\sqrt{\langle n_2\rangle(\langle n_2\rangle+1)}e^{i\left[(\phi_{M_1}-\phi_{M_2})-(\omega_1-\omega_2)t\right]}\label{M37}
\end{eqnarray}
Therefore, the sum of the phases of the forward and backward interference fringes only depends on the phase of the driving and the dynamical phase. Thus we have the phase constant $\phi_s = \phi_{\theta}+\phi_{\theta+\pi}= (\phi_{M_1}-\phi_{M_2})-(\omega_1-\omega_2)t$ and $g_+(\pi)=1$.

Meanwhile, phase correlation $g_{-}(\theta=\pi)$ is proportional to the mean value of $\hat{I}_f\hat{I}_b^\dagger$, together with Eq.~(\ref{eq:Iminus}) we have
\begin{eqnarray}
\langle e^{i\phi_{\theta}-\phi_{\theta+\pi}} \rangle \propto \langle \hat{I}_f\hat{I}_b^\dagger \rangle &=& \langle\psi^{(1)}|_S \otimes \langle\psi^{(2)}|_S\left(a_{\vk_1} a^\dagger_{\vk_2} a^\dagger_{-\vk_1} a_{-\vk_2}\right) |\psi^{(1)}\rangle_S \otimes |\psi^{(2)}\rangle_S \nonumber\\
&=&\langle\psi^{(1)}|_S\left( a_{\vk_1} a^\dagger_{-\vk_1}\right) |\psi^{(1)}\rangle_S \langle\psi^{(2)}|_S\left( a^\dagger_{\vk_2} a_{-\vk_2}\right) |\psi^{(2)}\rangle_S\nonumber\\
&=&  \langle\hat{I}_{1-}\rangle\langle \hat{I}_{2-}^\dagger\rangle\nonumber\\
&=& 0\label{M38}
\end{eqnarray}
therefore we have $g_{\_}(\pi)=0$, indicating that phases in each pair of jets are totally random although their sum is fixed. The results from Eqs.~(\ref{M37}, \ref{M38}) are consistent with our measurement shown in Fig.~3\textbf{g}.

Based on the same techniques of Eqs.~(\ref{M37}, \ref{M38}) and the methods in Ref.~\cite{Logan2017}, we derive a more general analytic formula for $g_+(\theta)$ and $g_-(\theta)$. We still use the symbol $\vk_1$ and $\vk_2$ corresponding two jets with different energy propagating along the same direction. We introduce $\vk^\prime_1$ and $\vk^\prime_2$ representing two jets propagating along the direction with a relative angle $\theta$ to $\vk_1$ and $\vk_2$. Therefore, $g_{\pm}(\theta)$ is written as
\begin{eqnarray}
g_+(\theta)=\left|{\langle a_{\vk_1}a^\dagger_{\vk_2} a_{\vk^\prime_1} a^\dagger_{\vk^\prime_2} \rangle \over \langle a^\dagger _{\vk_1} a_{\vk_1}\rangle \mean{a^\dagger_{\vk_2} a_{\vk _2} }}\right|=\left|{\langle a_{\vk_1}a_{\vk^\prime_1}\rangle \langle a^\dagger_{\vk_2} a^\dagger_{\vk^\prime_2} \rangle \over \langle a^\dagger _{\vk_1} a_{\vk_1}\rangle \mean{a^\dagger_{\vk_2} a_{\vk _2} }}\right|, \\
g_-(\theta)=\left|{\langle a_{\vk_1}a^\dagger_{\vk_2} a^\dagger_{\vk^\prime_1} a_{\vk^\prime_2} \rangle \over \langle a^\dagger _{\vk_1} a_{\vk_1}\rangle \mean{a^\dagger_{\vk_2} a_{\vk _2} }}\right|=\left|{\langle a_{\vk_1}a^\dagger_{\vk^\prime_1}\rangle \langle a^\dagger_{\vk_2} a_{\vk^\prime_2} \rangle \over \langle a^\dagger _{\vk_1} a_{\vk_1}\rangle \mean{a^\dagger_{\vk_2} a_{\vk _2} }}\right|.
\end{eqnarray}
According to the methods in Ref.~\cite{Logan2017}, we obtain
\begin{eqnarray}
    \mean{a_{\vk_j}a_{\vk^\prime_j}}&=&e^{i(\phi_{M_j}-\omega_jt-\pi/2)} {\tilde \rho(\vk_j+\vk^\prime_j) \over 2\pi}\cosh(\gamma_j)\sinh(\gamma_j) ,\\
    \mean{a^\dagger_{\vk_j}a_{\vk^\prime_j}}&= &{\tilde \rho(\vk_j-\vk^\prime_j) \over 2\pi}\sinh^2(\gamma_j),
\end{eqnarray}
where $\tilde\rho(\vk)$ is defined as the Fourier transformation of a uniform disk-shape density $\rho(\vvr)$,
\begin{equation}
    \rho(\vvr)={1\over 2\pi}\int d^2\vk e^{i\vk\cdot\vvr}\tilde \rho(\vk).
\end{equation}
And $\rho(\vvr)$ is the density distribution function of the condensate as
\begin{equation}
     \rho(\vvr)=\begin{cases}
1 & |\vvr|\leq R\\
0 & |\vvr|>R
\end{cases}
\end{equation}
with $R$ is the radius. Therefore, the analytic formulas for $g_{\pm}(\theta)$ when $\gamma_j\gg1$ and $|\vk_i|R\gg 1$ are
\begin{eqnarray}
g_+(\theta)&=&\left|{4\tilde\rho(\vk_1+\vk^\prime_1)\tilde\rho^*(\vk_2+\vk^\prime_2)\over \tilde R^4}\right| \nonumber \\
&=&\left|{4J_1(|\vk_1|R(\theta-\pi)) J_1(|\vk_2|R(\theta-\pi))\over|\vk_1||\vk_2| R^2(\theta-\pi)^2}\right| ,
\end{eqnarray}
and
\begin{eqnarray}
g_-(\theta)&=&\left|{4\tilde\rho(\vk_1-\vk^\prime_1)\tilde\rho^*(\vk_2-\vk^\prime_2)\over \tilde R^4}\right|\nonumber \\
&=&\left|{4J_1(|\vk_1|R\theta) J_1(|\vk_2|R\theta)\over |\vk_1||\vk_2| R^2\theta^2}\right|,
\end{eqnarray}
where $J_1(x)$ is the first order Bessel function of the first kind.

To experimentally extract the interference fringe phase $\phi_\theta$ at a particular emission direction $\theta$, we average over an angular span from $\theta-0.12$ to $\theta+0.12$ to obtain the radial density distribution $\rho(r,\theta)$ in order to achieve the best signal to noise ratio (see Fig.~\ref{fig3}\textbf{d}). We then perform Fourier transformation on the radial density to get the complex density amplitude of the interference fringes in momentum space $\rho(k,\theta)$. The phase $\phi_\theta$ at $k_f$ is then evaluated from this complex amplitude. Although our jet width is small, that is 2$^\circ$ for $\omega/2\pi$~=~3~kHz and 1.5$^\circ$ for $\omega/2\pi$~=~5.63~kHz, this average results in a significantly broadened phase correlation shown in Fig.~\ref{fig3}\textbf{g}. To experimentally extract the phase constant $\phi_s$, we fit the histogram of $\phi_\theta +\phi_{\theta+\pi}$ to get the peak position. We also calculate the expected phase shift based on our experimental sequence with a time of 18.5~ms from the start of the modulation to the start of imaging. The first sinusoidal modulation pulse lasts for 6 periods while the second lasts for 17 periods. Meanwhile we take into account the time delay of the modulation pulse of 0.041~ms due to system response. Therefore the phase constant estimated from our experimental sequence is 0.9(2) where the uncertainty arises from the duration of our 20~$\mu s$ imaging pulse.

\section{Numerical results on reversal of atomic radiation field}

In this section, we use numerical simulation based a dynamical Gross-Pitaevskii equation to investigate the partial reversal on radiating matter-wave fields. We find that this imperfect reversal results mostly come from the off-resonant coupling to finite momentum modes close to $|\mathbf{k}|=k_f$.

\begin{figure}
\begin{center}
\includegraphics[width=185mm]{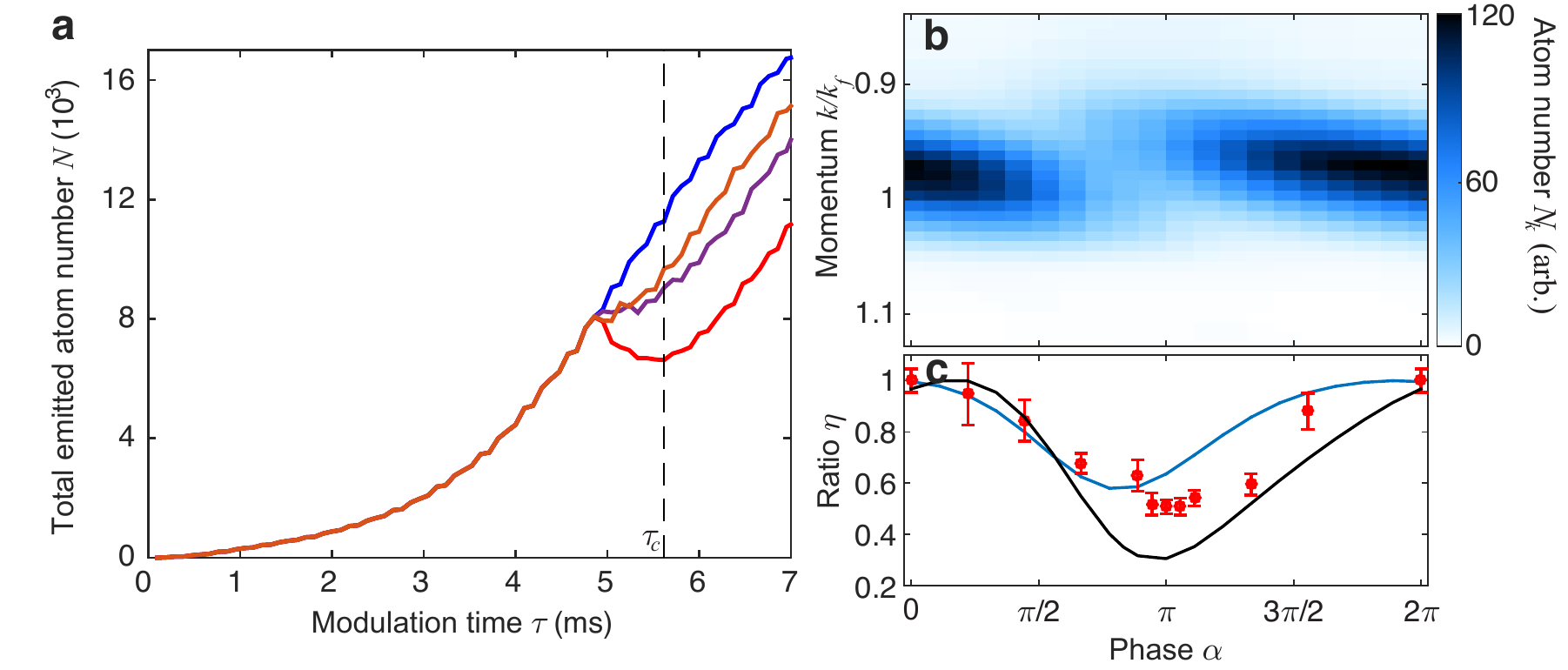}
\caption{\textbf{Numerical simulation on reversal of matter-wave radiation.} \textbf{a} shows the time evolution of the total emitted atom number $N$ for different phase jump $\alpha$~=~0 (blue), 0.44$\pi$ (purple), 0.89$\pi$ (red), and 1.33$\pi$ (orange). Here the phase jump happens at $\tau$~=~4.76~ms while optimal suppression is achieved at $\tau_c$~=~5.62~ms. \textbf{b} shows the emitted atom number at different momentum modes $N_k(\alpha)$ as a function of phase jump $\alpha$ at $\tau_c$.
\textbf{c} compares the overall suppression ratio $\eta$ from simulation (blue curve), suppression ratio for a particular momentum mode $k_f$ (black curve), and experiment (dots).}
\label{fig_sub_sim}
\end{center}
\end{figure}

Here we start with Gross-Pitaevskii equation,
\begin{equation}\label{eq:GP}
\begin{aligned}
i\hbar\frac{\partial\psi}{\partial t}=&\left[-\frac{\hbar^2}{2m}\nabla^2+V({\bf{r}})+U_{dc}|\psi|^2 -\mu\right]\psi \\
&+U_{ac}f(t)|\psi|^2\psi,
\end{aligned}
\end{equation}
where $\psi$ is the wavefunction and $\mu~=~2\pi\hbar\times 19$~Hz is the static chemical potential of the condensate,
$V(r)$ is the disk-shaped trapping potential as a function of radius $r$ with $V(r) = 2\pi\hbar\times 300$~Hz for $13.6~\mu$m$\geq r\geq13~\mu$m and $V(r)=0$ for the rest, $U_{dc}=4\pi\hbar^2a_{dc}/m$ and $U_{ac}=4\pi\hbar^2a_{ac}/m$ are
the DC and AC interaction strengths with
$a_{dc}~=~3a_0$ and $a_{ac}~=~50a_0$. In addition, we have $f(t) = \sin(\omega t)$ when $t\leq4.76$~ms and $f(t) = \sin(\omega t + \alpha)$ for $t>4.76$~ms. These parameters are chosen according to our experimental conditions.

The results from simulation using a CUDA-based solver \cite{GaugeFieldPaper} shows great agreement with the experiment. First of all, the total emitted atom number is suppressed after the phase jump $\alpha$ close to $\pi$ (see Fig.\ref{fig_sub_sim}\textbf{a}). The suppression sensitively depends on the phase of the second pulse $\alpha$. Similarly to the analysis of our experimental data, we then look at the suppression ratio as a function of phase $\alpha$ at $\tau_c~=~5.62$~ms when the optimal suppression appears (see Fig.\ref{fig_sub_sim}\textbf{c}). The suppression ratio varies as a function of $\alpha$ in the same way as in our experiment and the best suppression can be achieved is $\eta =0.57$ comparable to the experimental result.

The reason for this partial suppression is the off-resonant coupling to finite momentum modes close to $|\mathbf{k}|=k_f$. We examine more carefully about the emitted atoms in different momentum modes (see Fig.~\ref{fig_sub_sim}\textbf{b}), not all atoms are excited with a particular well-defined momentum. Instead, atoms spread across a range of momentum modes due to uncertainty principle since the atoms are confined within a finite radius of $13~\mu$m. These momentum modes are then off-resonantly coupled to the external modulation without perfect phase matching. Therefore population in these off-resonantly excited modes are maximally reversed at different phase jumps. For one particular momentum mode, the population can be reduced by as much as 70\%, while the overall population is only suppressed to about 50\%, consistent with our measurement.

Beside this off-resonant coupling, we anticipate the reversal can be limited by other effects such as the fast counter-rotating terms and the motion of the emitted atoms as well. The counter-rotating terms lead to quick population oscillations seen in Fig.~\ref{fig_sub_sim}\textbf{a}; they also accumulate phase and eventually limit the reversal. Furthermore, when atoms move out of the condensate, they can not be transferred back to the condensate anymore. These effects are included in the simulation but their contributions to the limited reversal are hard to separate in our numerical model.

\end{document}